\documentclass[sn-mathphys,pdflatex,Numbered]{sn-jnl}
\pdfoutput=1 

\newlength{\lw}
\usepackage{layout}
\usepackage{layouts}
\usepackage{gensymb}
\usepackage{circledsteps}
\newcommand{\CircledSmall}[2][]{%
\CircledParamOpts{inner ysep=3pt, inner xsep=0pt, #1}{1}{#2}%
}
\usepackage{siunitx}
\usepackage{empheq}
\usepackage[scaled=0.86]{helvet}
\usepackage{comment}
\usepackage{array}
\usepackage{bm}
\usepackage{csquotes}
\usepackage{cleveref}
\crefname{equation}{}{}
\Crefname{equation}{Equation}{Equations}
\crefname{figure}{Fig.}{Figs.}
\Crefname{figure}{Figure}{Figures}
\crefname{table}{Tab.}{Tabs.}
\Crefname{table}{Table}{Tables}
\crefname{section}{Sec.}{Secs.}
\Crefname{section}{Section}{Sections}
\crefname{algorithm}{Alg.}{Algs.}
\Crefname{algorithm}{Algorithm}{Algorithms}
\crefname{appendix}{}{}
\crefmultiformat{equation}{(#2#1#3)}%
{ and~(#2#1#3)}{, (#2#1#3)}{, and~(#2#1#3)}

\usepackage{graphicx}%
\usepackage{multirow}%
\usepackage{amsmath,amssymb,amsfonts,newtxtext}%
\usepackage{amsthm}%
\usepackage{mathrsfs}%
\usepackage[title]{appendix}%
\usepackage{xcolor}%
\usepackage{textcomp}%
\usepackage{manyfoot}%
\usepackage{booktabs}%
\usepackage{algorithm}%
\usepackage{algorithmicx}%
\usepackage{algpseudocode}%
\usepackage{listings}%
\makeatletter
\g@addto@macro\normalsize{%
  \setlength\abovedisplayskip{5pt}
  \setlength\belowdisplayskip{5pt}
  \setlength\abovedisplayshortskip{5pt}
  \setlength\belowdisplayshortskip{5pt}
}
\makeatother

\theoremstyle{thmstyleone}%
\theoremstyle{thmstyletwo}%

\theoremstyle{thmstylethree}%

\begin{document}
\title[Article Title]{A Unified Approach for Dynamic Analysis of Tensegrity Structures 
with Arbitrary Rigid Bodies and Rigid Bars}


\author[1]{\fnm{Jiahui} \sur{Luo}}\email{luojh66@mail.sysu.edu.cn}

\author*[1,2,3]{\fnm{Xiaoming} \sur{Xu}}\email{xuxm29@mail.sysu.edu.cn}

\author[1]{\fnm{Zhigang} \sur{Wu}}\email{wuzhigang@mail.sysu.edu.cn}

\author[1]{\fnm{Shunan} \sur{Wu}}\email{wushunan@mail.sysu.edu.cn}

\affil*[1]{
    \orgdiv{School of Aeronautics and Astronautics}, 
    \orgname{Shenzhen Campus of Sun Yat-sen University}, 
    \orgaddress{
        \street{No.66 Gongchang Road, Guangming District}, 
        \city{Shenzhen}, 
        \postcode{518107}, 
        \state{Guangdong}, 
        \country{P.R. China}
    }
}

\affil[2]{
    \orgdiv{Shenzhen Key Laboratory of Intelligent Microsatellite Constellation}, 
    \orgname{Shenzhen Campus of Sun Yat-sen University}, 
    \orgaddress{
        \street{No.66 Gongchang Road, Guangming District}, 
        \city{Shenzhen}, 
        \postcode{518107}, 
        \state{Guangdong}, 
        \country{P.R. China}
    }
}

\affil[3]{
    \orgdiv{State Key Laboratory of Structural Analysis for Industrial Equipment}, 
    \orgname{Dalian University of Technology}, 
    \orgaddress{
        \street{No.2 Linggong Road, Ganjingzi District}, 
        \city{Dalian}, 
        \postcode{116024}, 
        \state{Liaoning}, 
        \country{P.R. China}
    }
}



\abstract{This paper proposes a unified approach for dynamic modeling and
simulations of general tensegrity structures with rigid bars and rigid bodies of
arbitrary shapes. The natural coordinates are adopted as a non-minimal
description in terms of different combinations of basic points and base vectors 
to resolve the heterogeneity between rigid bodies and rigid bars in
three-dimensional space. This leads to a set of differential-algebraic equations
with a constant mass matrix and free from trigonometric functions. Formulations
for linearized dynamics are derived to enable modal analysis around static
equilibrium. For numerical analysis of nonlinear dynamics, we derive a modified
symplectic integration scheme which yields realistic results for long-time
simulations, and accommodates non-conservative forces as well as boundary
conditions. Numerical examples demonstrate the efficacy of the proposed approach
for dynamic simulations of Class-1-to-$k$ general tensegrity structures  
under complex situations, including dynamic external loads, cable-based
deployments, and moving boundaries. 
The novel tensegrity structures also exemplify new ways to create
multi-functional structures.
}




\maketitle
\section*{Article Highlights}
\begin{itemize}
	\item 
The	natural coordinates is reformed to develop a unified dynamic modeling approach for general tensegrity structures
	\item 
A modified symplectic integration (MSI) scheme is derived to solve the governing differential-algebraic equations	
	\item
Simulations for various general tensegrity structures under complex dynamic situations demonstrate the efficacy of MSI
\end{itemize}

\section{Introduction}
\label{sec:intro}
The term \textit{tensegrity}, combining \enquote{tensile} and
\enquote{integrity}, was coined by Buckminster Fuller
\cite{buckminsterTensileintegrityStructures1962} to describe a kind of
prestressed structures, that were created by Ioganson and Snelson
\cite{lalvaniOriginsTensegrityViews1996}. A commonly adopted definition is given
by Ref.  \cite{skeltonTensegritySystems2009}: a tensegrity structure is a
self-sustaining composition of rigid members and tensile members, and if there
is at least a torqueless joint connecting $k$ rigid members, it is called a
Class-$k$ tensegrity. Recent decades have witnessed two trends of developments
in the tensegrity literature.

One trend is interested in the so-called \enquote{bars-only} tensegrity
structures, where the rigid members are axial-loaded thin bars. This setting can
maximize material efficiency, making them strong and lightweight 
\cite{skeltonTensegritySystems2009}. 
Furthermore, they are found to be deployable using simple cable-based actuations
\cite{furuyaConceptDeployableTensegrity1992,sultanDeploymentTensegrityStructures2003,krishnanDesignLightweightDeployable2018}.
Thus, this trend is mostly seen in civil engineering
\cite{sultanChapterTensegrity602009,sychterzUsingDynamicMeasurements2018,veuveDeploymentTensegrityFootbridge2015}
and aerospace engineering
\cite{tibertDeployableTensegrityReflectors2002,chenDesignControlTensegrity2020,chenDesignAnalysisGrowable2020},
etc. 
The dynamic modeling and simulation problems of \enquote{bars-only} tensegrities 
were addressed by the non-minimal coordinates 
\cite{skeltonDynamicsControlTensegrity2005,skeltonEfficientModelsMultibody2010,
nagaseNetworkVectorForms2014,cheongNonminimalDynamicsGeneral2015,goyalTensegritySystemDynamics2019,
hsuLagrangianMethodConstrained2021} and quaternion-based formulations \cite{cefaloComprehensiveDynamicModel2011}. 
In particular, the non-minimal approach have advantages of a constant mass matrix and the absence of trigonometric functions.

The other trend concerns tensegrities with rigid bodies, which are allowed to
have complex shapes such as the \enquote{X-Piece}
\cite{snelsonArtTensegrity2012}. Compared to the \enquote{bars-only} setting,
they usually have simpler connectivity, larger capacity spaces, while still
being modular and compliant. Furthermore, they mimic the interactions of muscles
and bones \cite{skeltonTensegritySystems2009,sultanChapterTensegrity602009} such
as the vertebrate spine \cite{levinTENSEGRITYTRUSSMODELSPINE2002}, and thus often lead to bio-inspired designs,
including tensegrity joints \cite{lessardLightweightMultiaxisCompliant2016} and
tensegrity fishes \cite{chenSwimmingPerformanceTensegrity2019}, etc.
The dynamic problems of tensegrities with rigid bodies were previously addressed 
by incorporating tensile cables into the established multi-rigid-body dynamics
\cite{kanSlidingCableElement2018,kanComprehensiveFrameworkMultibody2021,mirletzBridgingRealityGap2015}.

In recent years, a growing interest in merging these two trends leads to the
so-called general tensegrity structures, which have the potential of combining
the above advantages.
For instance, Liu et al. \cite{liuKinematicStaticAnalysis2020} studied the
kinematics and statics of a fusiform tensegrity 
which combines a triangular rigid body and a rigid bar. Ma et al. formulated the
static equilibrium equations for form-finding problems of Class-1 general
tensegrities \cite{maEquilibriumFormFindingGeneral2022a}. Wang et al.
\cite{wangTopologyDesignGeneral2020} devised a topology-finding method to
discover new structures like the tensegrity bridge,
which has bars as supporting struts and a rigid plate as the bridge deck. They
also proposed a self-stress design method
\cite{wangSelfequilibriumMechanismStiffness2023} for Class-$k~(k\ge1)$ general
tensegrities, and provided illustrative two-dimensional examples. However, none
of these works address the dynamic simulation problems which are crucial in many
engineering applications, such as dynamic response analysis under external loads and 
dynamic deployment analysis.

To develop numerical methods for dynamic analysis of general tensegrities,
challenges arise from the heterogeneity between rigid bodies and rigid bars in
3D space. Firstly, the vanishing inertia about the longitudinal axis of a thin
bar can lead to singular mass matrices
\cite{wroldsenModellingControlNonminimal2009}. Secondly, the rotation and
angular velocity about the axis of a bar are ill-defined. In other words, a 3D rigid bar has only five degrees of freedom
(DoFs), one less than a 6-DoF 3D rigid body
\cite{cefaloComprehensiveDynamicModel2011}. Thirdly, a rigid bar can be jointed
only at its two endpoints, while a rigid body can be jointed at anywhere. 
In this paper, we address these problems by comprehensively reforming the natural coordinates 
\cite{xuGeneralizedInertiaRepresentation2021,dejalonNaturalCoordinatesComputer1986,
garciadejalonDynamicAnalysisThreeDimensional1987,dejalonTwentyfiveYearsNatural2007}. 
Specifically, both rigid bars and rigid bodies are nonminimally described by different 
combinations of basic points and base vectors, which form different types of natural coordinates. 
This nonminimal description resolves the above-mentioned singularity and ill-definedness problems, 
and the exhaustive types of coordinates facilitate sharing basic points for 
jointed rigid members.
By use of polymorphism and conversion matrices, 
these formulations are abstracted in succinct mathematical expressions, 
and thereby can be implemented as generic computer codes.
Additionally, the generalized tension forces of tensile cables are explicitly
derived, while boundary conditions are dealt with using a coordinate-separating strategy. 

This unified approach can not only model 3D Class-$k~(k\ge1)$ general tensegrity
structures, but also retain the advantages of non-minimal coordinates.
Nonetheless, it formulates the nonlinear dynamics in the form of
differential-algebraic equations (DAEs), where algebraic equations are present
to enforce the constraints for rigid members and joints. These constraints must
be satisfied in dynamic analyses to avoid degradation of accuracy. With this
consideration in mind, this paper develops solution methods for both constrained
linearized dynamics and constrained nonlinear dynamics. On the one hand, the
dynamics linearized around static equilibrium is reduced to the degrees of
freedom using the reduced-basis method, allowing accurate computations of
natural frequencies and mode shapes. On the other hand, a modified symplectic
integration (MSI) scheme is derived for numerical simulations of the constrained
nonlinear dynamics. The MSI scheme belongs to the Zu-class symplectic schemes
\cite{zhongIntegrationConstrainedDynamical2006,wuConstrainedHamiltonVariational2016}
that feature realistic behaviors in long-time simulations as well as exact
enforcement of algebraic constraints. Besides, this scheme is recast from the
viewpoint of approximations and limits in order to accommodate non-conservative
forces and boundary conditions. 

Finally, numerical examples are provided to validate the proposed modeling
approach and dynamic analysis methods. In particular, Class-$k~(k>1)$
tensegrities with jointed rigid bars and rigid bodies that are rarely seen in the
literature are presented. They demonstrate intuitive ways to design innovative
general tensegrities with potential multi-functionalities.

The rest of this paper is organized as follows. \cref{sec:bodies_and_bars}
derives the unified formulations for 3D rigid bodies and rigid bars,
based on which \cref{sec:tensegrity} model general tensegrity structures.
\cref{sec:dynamic} derives modal analysis and nonlinear dynamic analysis
methods, followed by the numerical examples in \cref{sec:examples}. Finally,
conclusions are drawn in \cref{sec:conclusion}.

\section{Unifying rigid bodies and rigid bars using natural
coordinates}\label{sec:bodies_and_bars} In this section, the natural coordinates
\cite{dejalonKinematicDynamicSimulation1994,pappalardoNaturalAbsoluteCoordinate2015}
are adapted for unifying the non-minimal descriptions of rigid bodies and rigid
bars, which are collectively called rigid members, and indistinguishably labeled
by circled numbers \CircledSmall{1}, \CircledSmall{2}, \dots, or circled capital
letters \CircledSmall{$I$}, \CircledSmall{$J$}, \dots, etc. Thus, a quantity
with a capital subscript, such as $()_{I}$, indicates the quantity belongs to
the $I$th rigid member.

\subsection{Rigid bodies of arbitrary shapes}\label{sec:rigidbodies}
\subsubsection{3D rigid bodies}\label{sec:3d_rigid}
\begin{figure}[tbh]
	\centering
	\includegraphics[width=251pt]{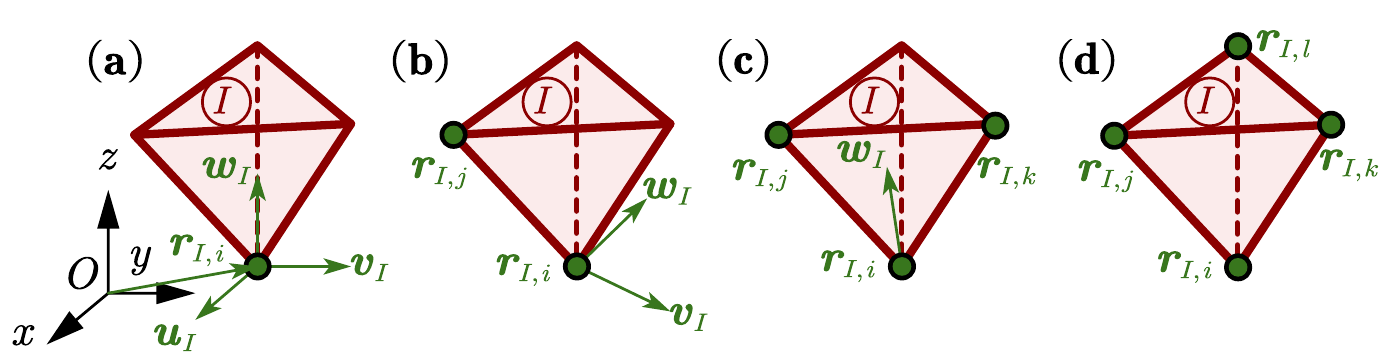}
	\caption{A 3D rigid body described by four types of natural coordinates. Rigid bodies are drawn by red lines. Basic points and base vectors are colored in green.}
	\label{fig:3D_NC}
\end{figure}
Consider a tetrahedron which exemplifies an arbitrary 3D rigid body, as shown in
\cref{fig:3D_NC}, where basic points
$\bm{r}_{I,i},\bm{r}_{I,j},\bm{r}_{I,k},\bm{r}_{I,l}\in \mathbb{R}^3$ and base
vectors $\bm{u}_I,\bm{v}_I,\bm{w}_I\in \mathbb{R}^3$ are fixed on the rigid body
and expressed in the global inertial frame $Oxyz$. Four types of natural
coordinates, i.e.
\begin{equation}\label{eq:3D_natural}
	\begin{aligned}
		\bm{q}_{I,\mathrm{ruvw}}&=[\bm{r}^\mathrm{T}_{I,i},\bm{u}_I^\mathrm{T},\bm{v}_I^\mathrm{T},\bm{w}_I^\mathrm{T}]^\mathrm{T},\ \  \bm{q}_{I,\mathrm{rrvw}}=[\bm{r}^\mathrm{T}_{I,i},\bm{r}^\mathrm{T}_{I,j},\bm{v}_I^\mathrm{T},\bm{w}_I^\mathrm{T}]^\mathrm{T},\\
		\bm{q}_{I,\mathrm{rrrw}}&=[\bm{r}^\mathrm{T}_{I,i},\bm{r}^\mathrm{T}_{I,j},\bm{r}^\mathrm{T}_{I,k},\bm{w}_I^\mathrm{T}]^\mathrm{T},\ \  \mathrm{and}\ \ 
		\bm{q}_{I,\mathrm{rrrr}}=[\bm{r}^\mathrm{T}_{I,i},\bm{r}^\mathrm{T}_{I,j},\bm{r}^\mathrm{T}_{I,k},\bm{r}^\mathrm{T}_{I,l}]^\mathrm{T}\in \mathbb{R}^{12},
	\end{aligned}
\end{equation}
can be used to describe a 3D rigid body, corresponding to \cref{fig:3D_NC} (a)
to (d), respectively, where $()_\mathrm{ruvw}$, etc, denote the type of natural
coordinates. For the latter three types of natural coordinates, we can formally
define $\bm{u}_I=\bm{r}_{I,j}{-}\bm{r}_{I,i}$,
$\bm{v}_I=\bm{r}_{I,k}{-}\bm{r}_{I,i}$, and
$\bm{w}_I=\bm{r}_{I,l}{-}\bm{r}_{I,i}$, so that they can be converted to the
first type by 
\begin{equation}\label{eq:3D_conversion}
	\bm{q}_{I,\mathrm{ruvw}}=
	\bm{Y}_{\mathrm{ruvw}}\bm{q}_{I,\mathrm{ruvw}}=
	\bm{Y}_{\mathrm{rrvw}}\bm{q}_{I,\mathrm{rrvw}}=
	\bm{Y}_{\mathrm{rrrw}}\bm{q}_{I,\mathrm{rrrw}}=
	\bm{Y}_{\mathrm{rrrr}}\bm{q}_{I,\mathrm{rrrr}},
\end{equation}
where the conversion matrices are defined as, respectively,
\begin{equation}\label{key}
	\begin{aligned}
		\bm{Y}_{\mathrm{ruvw}}&=\begin{bsmallmatrix*}[r]
			1& 0& 0& 0\\
			0& 1& 0& 0\\
			0& 0& 1& 0\\
			0& 0& 0& 1\\
		\end{bsmallmatrix*}\otimes\mathbf{I}_3,\,
		\bm{Y}_{\mathrm{rrvw}}=\begin{bsmallmatrix*}[r]
			1& 0& 0& 0\\
			-1& 1& 0& 0\\
			0& 0& 1& 0\\
			0& 0& 0& 1\\
		\end{bsmallmatrix*}\otimes\mathbf{I}_3,\, \\
		\bm{Y}_{\mathrm{rrrw}}&=\begin{bsmallmatrix*}[r]
			1& 0& 0& 0\\
			-1& 1& 0& 0\\
			-1& 0& 1& 0\\
			0& 0& 0& 1\\
		\end{bsmallmatrix*}\otimes\mathbf{I}_3,\, \mathrm{and}\,
		\bm{Y}_{\mathrm{rrrr}}=\begin{bsmallmatrix*}[r]
			1& 0& 0& 0\\
			-1& 1& 0& 0\\
			-1& 0& 1& 0\\
			-1& 0& 0& 1\\
		\end{bsmallmatrix*}\otimes\mathbf{I}_3
	\end{aligned}
\end{equation}
where $\mathbf{I}_3$ is a $3\times3$ identity matrix, and $\otimes$ denotes the
Kronecker product.

Note that the base vectors are assumed to be non-coplanar, thus the natural
coordinates actually form an affine frame attached to the 3D rigid body.
Consequently, the position vector of a generic point on the 3D rigid body can be
expressed by
\begin{equation}\label{eq:3D_trans}
	\bm{r}=\bm{r}_{I,i}+c_{I,1}\bm{u}_I+c_{I,2}\bm{v}_I+c_{I,3}\bm{w}_I=
	\bm{C}_{I,\mathrm{body}}\bm{q}_{I,\mathrm{body}},
\end{equation}
where $c_{I,1}$, $c_{I,2}$ and $c_{I,3}$ are the affine coordinates;
$\bm{C}_{I,\mathrm{body}}=\left(\left[1, c_{I,1},
c_{I,2}, c_{I,3}\right] \otimes \mathbf{I}_3\right)\bm{Y}_{\mathrm{body}}$ is a transformation matrix for
$\bm{q}_{I,\mathrm{body}}$; 
$()_{\mathrm{body}}$ can be any of 
$()_{\mathrm{ruvw}}$, $()_{\mathrm{rrvw}}$, $()_{\mathrm{rrrw}}$, or $()_{\mathrm{rrrr}}$.

To ensure rigidity of the body, the natural coordinates
$\bm{q}_{I,\mathrm{body}}$ must satisfy six intrinsic constraints
\begin{equation}\label{eq:3D_intrinsic}
	\begin{aligned}
		\bm{\varPhi}_I( \bm{q}_{I,\mathrm{body}} ) =
		\begin{psmallmatrix}	
			\bm{u}_I^\mathrm{T}\bm{u}_I-\bar{\bm{u}}_I^\mathrm{T}\bar{\bm{u}}_I\\	
			\bm{v}_I^\mathrm{T}\bm{v}_I-\bar{\bm{v}}_I^\mathrm{T}\bar{\bm{v}}_I\\	
			\bm{w}_I^\mathrm{T}\bm{w}_I-\bar{\bm{w}}_I^\mathrm{T}\bar{\bm{w}}_I\\	
			\bm{v}_I^\mathrm{T}\bm{w}_I-\bar{\bm{v}}_I^\mathrm{T}\bar{\bm{w}}_I\\	
			\bm{u}_I^\mathrm{T}\bm{w}_I-\bar{\bm{u}}_I^\mathrm{T}\bar{\bm{w}}_I\\	
			\bm{u}_I^\mathrm{T}\bm{v}_I-\bar{\bm{u}}_I^\mathrm{T}\bar{\bm{v}}_I\\
		\end{psmallmatrix}=\bm{0}
	\end{aligned}	
\end{equation}
where $\bar{\bm{u}}_I$, $\bar{\bm{v}}_I$ and $\bar{\bm{w}}_I$ are constant
vectors in a local frame, which is fixed on the rigid member (See also
\cref{sec:uni_mass}). Then, the position and orientation of a 6-DoF 3D rigid body
can be defined by twelve coordinates (any type in \cref{eq:3D_natural}) and six
constraints \cref{eq:3D_intrinsic}.
\subsection{3D rigid bars}\label{sec:bar}
\begin{figure}[tbh]
	\centering
	\includegraphics[width=162pt]{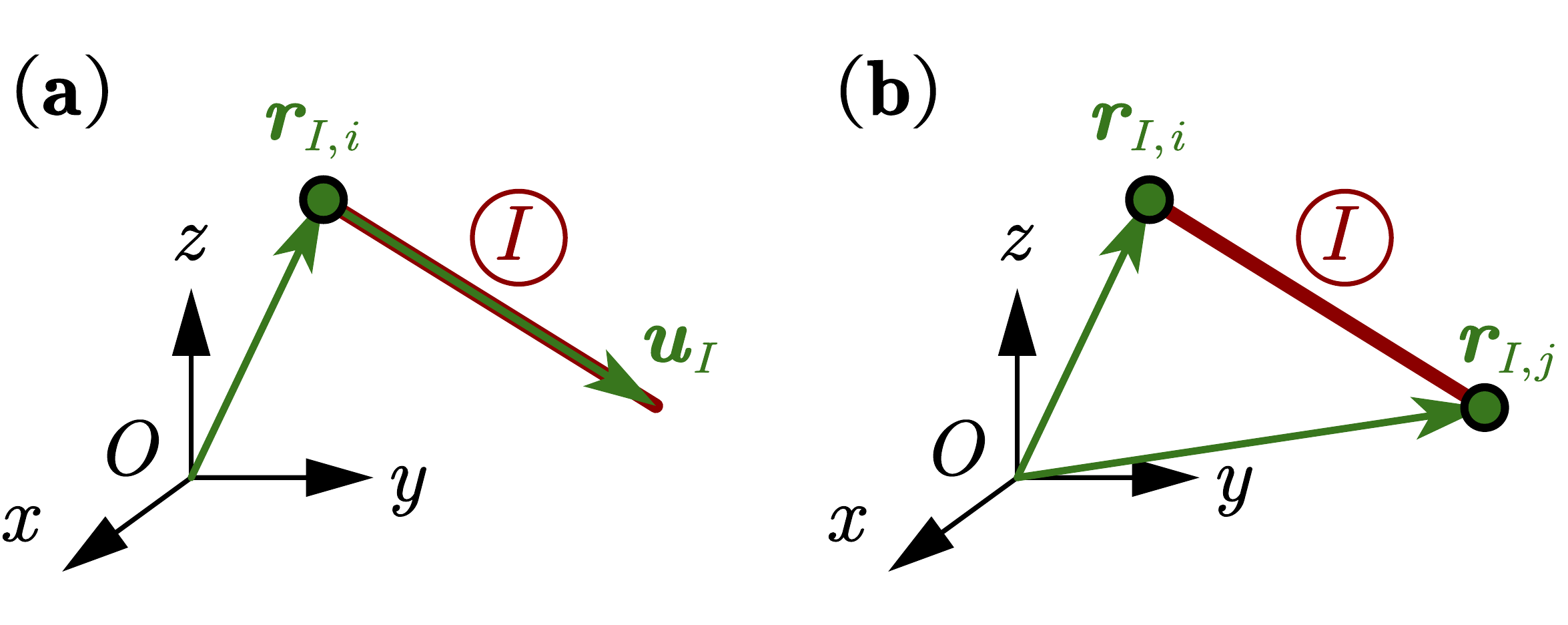}
	\caption{A 3D rigid bar described by two types of natural coordinates.}
	\label{fig:Bar}
\end{figure}
Two types of natural coordinates, i.e. 
$\bm{q}_{I,\mathrm{ru}}=[\bm{r}^\mathrm{T}_{I,i},\bm{u}_I^\mathrm{T}]^\mathrm{T}$
and 
$\bm{q}_{I,\mathrm{rr}}=[\bm{r}_{I,i}^\mathrm{T},\bm{r}_{I,j}^\mathrm{T}]^\mathrm{T}\in\mathbb{R}^{6}$,
can describe a 3D rigid bar, corresponding to \cref{fig:Bar} (a) and (b), respectively.
Define conversion matrices 
\begin{equation}	
	\bm{Y}_\mathrm{ru}=\begin{bsmallmatrix*}[r]
		1& 0\\
		0& 1\\
	\end{bsmallmatrix*}\otimes \bm{\mathrm{I}}_3\ \text{and}\ 
	\bm{Y}_\mathrm{rr}=\begin{bsmallmatrix*}[r]
		1& 0\\
		-1& 1\\
	\end{bsmallmatrix*}\otimes \bm{\mathrm{I}}_3
\end{equation}
Then, the position vector of a generic point along the longitudinal axis of the rigid
bar is given by
\begin{equation}\label{eq:bar_trans}
	\bm{r}=\bm{r}_{I,i}+c_I\bm{u}_I=\bm{C}_{I,\mathrm{bar}}\bm{q}_{I,\mathrm{bar}}
\end{equation}
where the coefficient $c_I$ depends on the relative position of the generic point;
$\bm{C}_{I,bar}=\left(\left[1, c_I\right] \otimes \mathbf{I}_3\right)\bm{Y}_{\mathrm{bar}}$ is the
transformation matrix for $\bm{q}_{I,\mathrm{bar}}$; 
$()_{\mathrm{bar}}$ can be either $()_{\mathrm{ru}}$ or $()_{\mathrm{rr}}$.
And the intrinsic constraint to preserve the bar length is
\begin{equation}\label{eq:bar_intrinsic}
	\varPhi_I( \bm{q}_{I,\mathrm{bar}} )=\bm{u}_I^\mathrm{T}\bm{u}_I-\bar{\bm{u}}_I^\mathrm{T}\bar{\bm{u}}_I=0
\end{equation}

Hence, the position and orientation of a 5-DoF 3D rigid bar can be
defined by six coordinates and one constraint \cref{eq:bar_intrinsic}.
\subsection{Unified formulations and mass matrices}\label{sec:uni_mass}

\begin{table}[tbh]
	\caption{Polymorphism of natural coordinates for rigid bodies and rigid bars}
	\label{tab:polymorphsim}
	\centering
	\setlength\tabcolsep{4pt}
	\renewcommand{\arraystretch}{1.0}
	\begin{tabular}{c|cccc}
	\toprule
	&
	{\shortstack{Degrees of \\ freedom}} & 
	{\shortstack{Number of \\ coordinates}} & 
	{\shortstack{Number of \\ constraints}} & 
	{\shortstack{Types of \\ natural coordinates}} \\
	\midrule
	3D Rigid Body & 6 & 12 & 6 & ruvw rrvw rrrw rrrr \\
	3D Rigid Bar  & 5 & 6  & 1 & ru rr \\
	\botrule
	\end{tabular}
\end{table}
The transformation relations \cref{eq:bar_trans,eq:3D_trans} for the
standard types of natural coordinates can be put into a unifying form
\begin{equation}\label{eq:trans}
	\bm{r}=\bm{C}_{I}\bm{q}_{I},
\end{equation}
which is a polymorphic expression, meaning that the formulations of
$\bm{C}_I$ and $\bm{Y}_I$ vary with the type of $\bm{q}_I$, as summarized in
\cref{tab:polymorphsim}. However, note that $\bm{C}_{I}$ is not a function of
$\bm{q}_I$. Consequently, the velocity of a generic point is given by
$\dot{\bm{r}}=\bm{C}_{I}\dot{\bm{q}}_{I}$, which can be used to derive the mass
matrix. Let $\rho_I$ denote the longitudinal or volume density of the rigid
member $\CircledSmall{I}$. Then, the kinetic energy can be computed by an
integral over its entire domain $\Omega$ as
\begin{equation}\label{eq:T_I}\textstyle
	\begin{aligned}\textstyle
		T_I
		&=\frac{1}{2}{\textstyle\int\limits_{\Omega}}{\rho_I\bm{\dot{r}}^\mathrm{T}\bm{\dot{r}}\mathrm{d}\Omega}
		=\frac{1}{2}{\textstyle\int\limits_{\Omega}}{\rho_I \dot{\bm{q}}_{I}^\mathrm{T}\bm{C}_{I}^\mathrm{T}\bm{C}_I\dot{\bm{q}}_I\mathrm{d}\Omega}=\frac{1}{2}\dot{\bm{q}}_{I}^\mathrm{T}\bm{M}_I\dot{\bm{q}}_I
	\end{aligned}
\end{equation}
where $\bm{M}_I$ is a constant mass matrix with polymorphism defined by
\begin{equation}\label{eq:M}
\begin{aligned}
	\bm{M}_{I}&
	=\int _{\Omega}\rho _I\bm{C}_{I}^{\mathrm{T}}\bm{C}_{I}\mathrm{d}\Omega
	=\bm{Y}_I^{\mathrm{T}}\left(\int _{\Omega}\left( \rho _I \begin{bmatrix}
		1&		\bm{c}_{I}^{\mathrm{T}}\\
		\bm{c}_I&		\bm{c}_I\bm{c}_{I}^{\mathrm{T}}
	\end{bmatrix} \right) \mathrm{d}\Omega \otimes \mathbf{I}_3\right)\bm{Y}_I\\
	&=\bm{Y}_I^{\mathrm{T}}\left(\begin{bmatrix}
		\smallint _{\Omega}\rho _I\mathrm{d}\Omega&		\smallint _{\Omega}\rho _I\bm{c}_{I}^{\mathrm{T}}\mathrm{d}\Omega\\
		\smallint _{\Omega}\rho _I\bm{c}_I\mathrm{d}\Omega&		\smallint _{\Omega}\rho _I\bm{c}_I\bm{c}_{I}^{\mathrm{T}}\mathrm{d}\Omega\\
	\end{bmatrix} \otimes \mathbf{I}_3\right)\bm{Y}_I
\end{aligned}
\end{equation}

\begin{figure}[tbh]
	\centering
	\includegraphics[width=244pt]{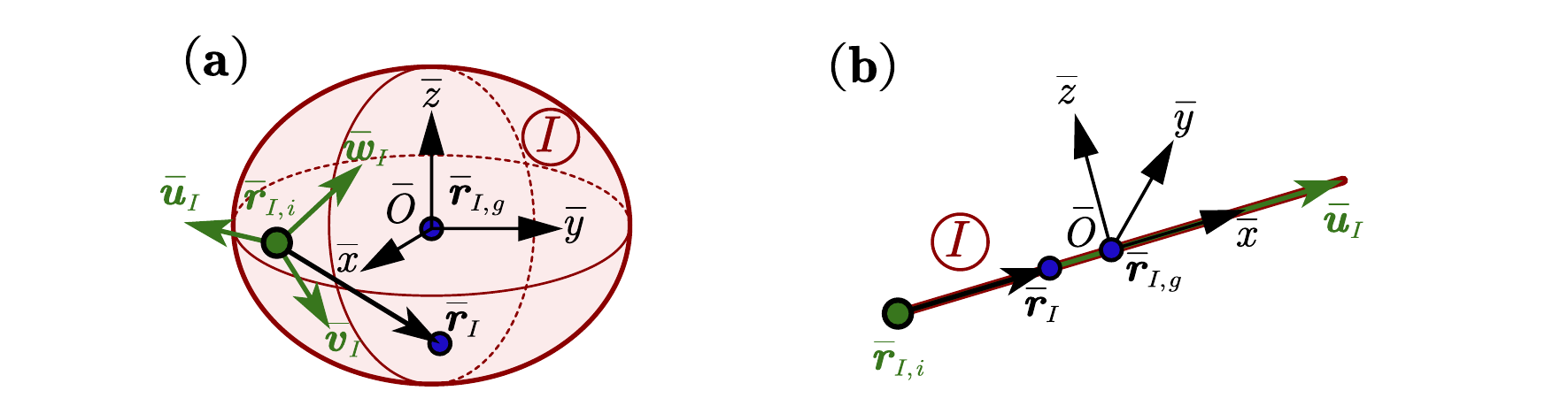}
	\caption{The basic point $\bar{\bm{r}}_{I,i}$, 
	the base vectors $\bar{\bm{u}}_{I}$, $\bar{\bm{v}}_{I}$ and $\bar{\bm{w}}_{I}$, 
	the mass center $\bar{\bm{r}}_{I,g}$, 
	and a generic point $\bar{\bm{r}}_{I}$ in the local Cartesian frame of 
	(a) a 3D rigid body or (b) a 3D rigid bar.}\label{fig:local}
\end{figure}

It is possible to express the mass matrix by conventional inertia properties, 
such as the mass, the center of mass, 
and the moments of inertia of a rigid member. 
To this end, let's introduce a local Cartesian frame $\bar{O}\bar{x}\bar{y}\bar{z}$
which is fixed on the rigid member $\CircledSmall{I}$, as shown in \cref{fig:local}. 
Quantities expressed in this local frame are denoted by an overline $\bar{()}$.
Without loss of generality, let its origin $\bar{O}$ coincide with the mass center, 
such that $\bar{\bm{r}}_{I,g}=\bm{0}$. 
For a 3D rigid body, let its axes align along the principal axes of inertia. 
For a 3D rigid bar, let its $\bar{x}$ axis aligns along the longitudinal direction.

Because the basic points and base vectors are fixed on the rigid members, their
coordinates in the local frame are constant. Let's define a polymorphic matrix
\begin{subequations}\label{eq:X}
	\begin{empheq}[left={\bar{\bm{X}}_I=\empheqlbrace}]{align}
		&\left[\bar{\bm{u}},\bar{\bm{v}},\bar{\bm{w}}\right] &&\mathrm{for~a~rigid~body}\label{eq:X_uvw}\\
		&\left[\bar{\bm{u}}\right] &&\mathrm{for~a~rigid~bar}\label{eq:X_u}
	\end{empheq}
\end{subequations}
Then, according to \cref{eq:trans}, the position vector of a generic point 
in the local frame can be expressed by
$\bar{\bm{r}} = \bar{\bm{r}}_{I,i} + \bar{\bm{X}}_I \bm{c}_I$, which gives
\begin{equation}\label{eq:c}
	\bm{c}_I=\bar{\bm{X}}^{+}_I( \bar{\bm{r}}-\bar{\bm{r}}_{I,i} )
\end{equation}
where $()^{+}$ denotes the Moore-Penrose pseudoinverse. For \cref{eq:X_uvw}, because the columns are linearly independent, i.e. $\bar{\bm{X}}$ has full rank, the pseudoinverse is equal to the matrix inverse.

Using \cref{eq:c}, the following expressions for use in \cref{eq:M} can be derived:
\begin{subequations}\label{eq:int_over}
	\begin{empheq}{align}
		\int _{\Omega}\rho _I\mathrm{d}\Omega &=m_I\\
		\int _{\Omega}\rho _I\bm{c}_I\mathrm{d}\Omega &= m_I\bar{\bm{X}}^+\left( \bar{\bm{r}}_{I,g}-\bar{\bm{r}}_{I,i} \right)=-m_I\bar{\bm{X}}^+\bar{\bm{r}}_{I,i}\\
		\int _{\Omega}\rho _I\bm{c}_I\bm{c}_{I}^{\mathrm{T}}\mathrm{d}\Omega &=
		\bar{\bm{X}}^+\left( \bar{\bm{J}}_I -m_I\bar{\bm{r}}_{I,i}\bar{\bm{r}}_{I,g}^{\mathrm{T}}-m_I\bar{\bm{r}}_{I,g}\bar{\bm{r}}_{I,i}^{\mathrm{T}}+m_I\bar{\bm{r}}_{I,i}\bar{\bm{r}}_{I,i}^{\mathrm{T}} \right) \bar{\bm{X}}^{+\mathrm{T}}\\
		&=\bar{\bm{X}}^+\left( \bar{\bm{J}}_I +m_I\bar{\bm{r}}_{I,i}\bar{\bm{r}}_{I,i}^{\mathrm{T}} \right) \bar{\bm{X}}^{+\mathrm{T}}
	\end{empheq}
\end{subequations}
where $m_I$ is the mass of the rigid member $\CircledSmall{I}$; 
$\bar{\bm{J}}_I$ contains the moments of inertia and necessitates some discussions:

For a 3D rigid body, 
$\bar{\bm{J}}_I$ is given by 
\begin{equation}\label{eq:J_3D}
	\bar{\bm{J}}_I 
	=\int_{\Omega}\rho_I\bar{\bm{r}}\bar{\bm{r}}^{\mathrm{T}}\mathrm{d}\Omega
	=\int_{\Omega}\rho_I\begin{bmatrix}
		\bar{x}^2     & \bar{y}\bar{x}& \bar{z}\bar{x}\\
		\bar{x}\bar{y}&	     \bar{y}^2& \bar{z}\bar{y}\\
		\bar{x}\bar{z}& \bar{y}\bar{z}&      \bar{z}^2\\
	\end{bmatrix} \mathrm{d}\Omega,
\end{equation}
while the conventional inertia matrix is given by 
\begin{equation}
	\bar{\bm{I}}_I=\int_{\Omega}\rho_I\begin{bmatrix}
		\bar{y}^2+\bar{z}^2     & -\bar{y}\bar{x}& -\bar{z}\bar{x}\\
		-\bar{x}\bar{y}&	     \bar{x}^2+\bar{z}^2& -\bar{z}\bar{y}\\
		-\bar{x}\bar{z}& -\bar{y}\bar{z}&      \bar{x}^2+\bar{y}^2\\
	\end{bmatrix} \mathrm{d}\Omega
\end{equation} 
Hence, we have $\bar{\bm{J}}_I =\tfrac{1}{2}\mathrm{trace}\left( \bar{\bm{I}}_I \right) \mathbf{I}_3-\bar{\bm{I}}_I$.

For a 3D rigid bar, the expression of $\bar{\bm{J}}_I$ is the same as \cref{eq:J_3D}, except that only the element $\bar{x}^2$ is nonzero. And the pseudoinverse of $\bar{\bm{X}}_{I}=\left[ \bar{u}_x,0,0 \right] ^{\mathrm{T}}$ is $\bar{\bm{X}}_{I}^{+}=\left[{1}/{\bar{u}_x},0,0\right]$. Therefore, we have $\bar{\bm{X}}_{I}^{+}\bar{\bm{J}}_I\bar{\bm{X}}_{I}^{+\mathrm{T}}=\left(\int_{\Omega}\rho_I\bar{x}^2\mathrm{d}\Omega\right)/{\bar{u}^2_x}$.

For an advanced treatment of the inertia representation for rigid multibody systems in terms of natural coordinates, 
we refer the interested readers to our previous paper \cite{xuGeneralizedInertiaRepresentation2021}.

\section{Modeling tensegrity structures}\label{sec:tensegrity}
Given the formulations of rigid members, 
the modeling of general Class-$k$ $(k\ge1)$ tensegrity structures additionally requires formulations for tensile cables, 
torqueless joints, and boundary conditions, which are derived in this section. 
A system is assumed to have $n_b$ rigid members and $n_s$ tensile cables.
\subsection{Ball joints}\label{sec:PJ}
\begin{figure}[tbh]
	\centering
	\includegraphics[width=258pt]{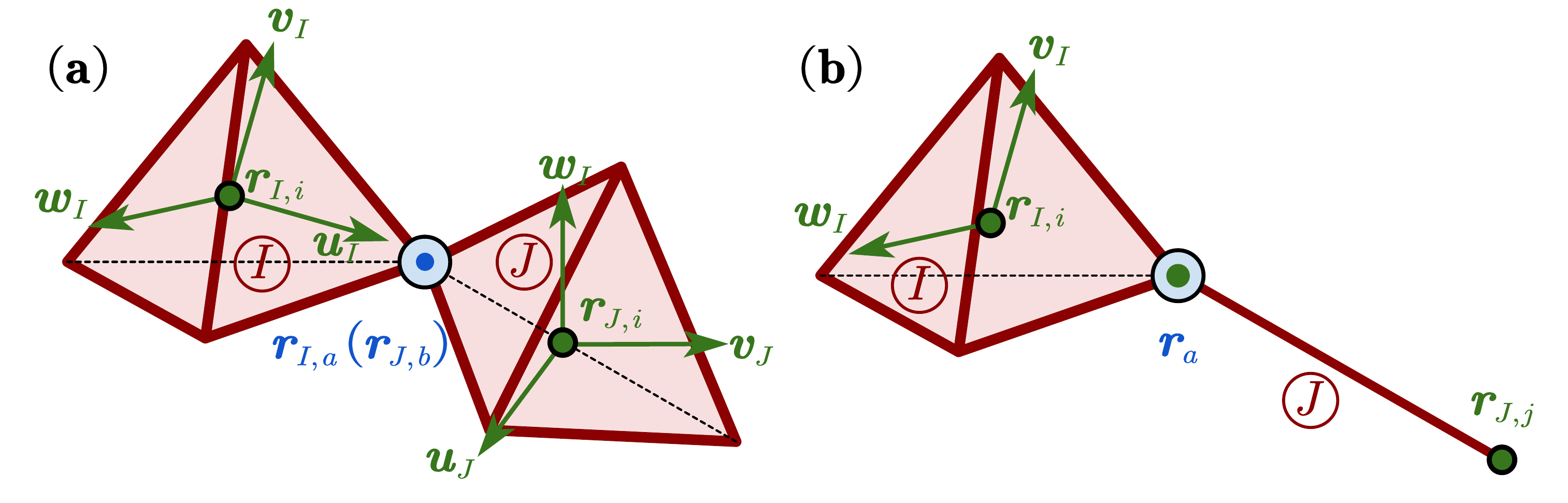}
	\caption{(a) Two 3D rigid bodies or (b) a 3D rigid body and a 3D rigid bar connected by a ball joint, which is represented by a circle filled with light blue.}
	\label{fig:PJ}
\end{figure}
A Class-$k~(k > 1)$ tensegrity structure allows the use of torqueless ball joints, 
each of which can connect up to $k$ different rigid members. 
Depending on the placements of basic points, there are two modeling methods. 

The first is a general method, as exemplified by \cref{fig:PJ} (a), where a ball joint connects point $a$ of rigid body $\CircledSmall{I}$ on point $b$ of rigid body $\CircledSmall{J}$, and consequently imposing a set of extrinsic constraints 
\begin{equation}\label{eq:ex_con}
	\bm{\varPhi}^\mathrm{ex}(\bm{q}_I,\bm{q}_J)= \bm{r}_{I,a} - \bm{r}_{J,b}=	\bm{C}_{I,a}\bm{q}_I-\bm{C}_{J,b}\bm{q}_J=\bm{0},
\end{equation}
where \cref{eq:trans} is used for the second equality.

The second method is to share the basic points between rigid members, 
as exemplified by \cref{fig:PJ} (b), 
where a ball joint is located at the basic point $a$. 
So we have natural coordinates 
$\bm{q}_I=[\bm{r}^\mathrm{T}_{I,i},\bm{r}^\mathrm{T}_{a},\bm{v}^\mathrm{T}_I,\bm{w}^\mathrm{T}_I]^\mathrm{T}$ 
for the rigid body $\CircledSmall{I}$, and $\bm{q}_J=[\bm{r}^\mathrm{T}_a,\bm{r}^\mathrm{T}_{J,j}]^\mathrm{T}$ 
for the rigid bar $\CircledSmall{J}$: they share the basic point's vector $\bm{r}_a$.  

If a ball joint connects $k$ ($k{>}2$) rigid members, 
it can be modeled as $k{-}1$ ball joints overlapping at one place. 

The second method has computational advantages over 
the first one because it needs no extrinsic constraint, 
and it reduces the number of system's coordinates.
Thanks to the exhaustion in deriving different combinations of 
the natural coordinates (\cref{sec:rigidbodies,sec:bar}), 
up to four or two basic points of a 3D rigid body or rigid bar 
can be used for sharing with other rigid members. 
Therefore, the second method is generally sufficient to model most Class-$k$ $(k{>}1)$ tensegrities, 
and the extrinsic constraints \cref{eq:ex_con} are rarely needed.

\subsection{Boundary conditions}\label{sec:prescribed}
In practice, most tensegrity structures have some members with prescribed motions, 
such that their positions, velocities, 
and accelerations are either partly or entirely given. 
For example, some rigid members in geodesic tensegrity domes are pin-jointed to the ground, 
or the rigid body motions of a self-standing tensegrity structure are to be eliminated. 
It would be cumbersome to derive case-by-case formulations 
for these prescribed rigid members. 
Alternatively, we can extend the above derivations, 
but also without loss of flexibility, 
by separating the prescribed and free (unprescribed) coordinates. 
To do this, let's denote the numbers of prescribed, free, 
and total coordinates for the rigid member $\CircledSmall{I}$ 
by $\tilde{n}_I$, $\check{n}_I$, and $n_I=\tilde{n}_I+\check{n}_I$, respectively, 
and for the system by $\tilde{n}$, $\check{n}$, and $n=\tilde{n}+\check{n}$, respectively. 
Then, the separation and reintegration of the coordinates of the rigid member $\CircledSmall{I}$ 
and of the system are defined by
\begin{equation}\label{eq:separate}
\begin{aligned}    
	\begin{pmatrix}\tilde{\bm{q}}_I\\\check{\bm{q}}_I\end{pmatrix}&=
	\begin{bmatrix}\tilde{\bm{E}}_I^\mathrm{T}\\\check{\bm{E}}_I^\mathrm{T}\end{bmatrix}\bm{q}_I,\ \ 
	\bm{q}_I=
	\begin{bmatrix}\tilde{\bm{E}}_I,\check{\bm{E}}_I\end{bmatrix}
	\begin{pmatrix}\tilde{\bm{q}}_I\\\check{\bm{q}}_I\end{pmatrix},\\
	\begin{pmatrix}\tilde{\bm{q}}\\\check{\bm{q}}\end{pmatrix}&=
	\begin{bmatrix}\tilde{\bm{E}}^\mathrm{T}\\\check{\bm{E}}^\mathrm{T}\end{bmatrix}\bm{q},\ \ \mathrm{and}\ \ 
	\bm{q}=\begin{bmatrix}\tilde{\bm{E}},\check{\bm{E}}\end{bmatrix}
	\begin{pmatrix}\tilde{\bm{q}}\\\check{\bm{q}}\end{pmatrix},
\end{aligned}
\end{equation}
where $\tilde{\bm{q}}_I\in\mathbb{R}^{\tilde{n}_I}$ and $\tilde{\bm{q}}\in\mathbb{R}^{\tilde{n}}$ are prescribed coordinates; 
$\check{\bm{q}}_I\in\mathbb{R}^{\check{n}_I}$ and $\check{\bm{q}}\in\mathbb{R}^{\check{n}}$ are free coordinates; 
$[\tilde{\bm{E}}_I,\check{\bm{E}}_I]\in\mathbb{Z}^{n_I\times n_I}$ and 
$[\tilde{\bm{E}},\check{\bm{E}}]\in\mathbb{Z}^{n\times n}$ are constant orthonormal matrices that only have zeros and ones as elements. 

The relations between the system's coordinates and those of rigid members and prescribed points are given by
\begin{equation}\label{eq:select}
	\bm{q}_I=\bm{T}_I\bm{q}=\tilde{\bm{T}}_I\tilde{\bm{q}}+\check{\bm{T}}_I\check{\bm{q}},\ \ 
	\mathrm{for}\ \ I=1,\dots,n_b,		
\end{equation}
where $\bm{T}_I$, $\tilde{\bm{T}}_I=\bm{T}_I\tilde{\bm{E}}$, and $\check{\bm{T}}_I=\bm{T}_I\check{\bm{E}}$ are constant matrices that select the right elements from the system, and also properly embody the sharing of basic points as presented in \cref{sec:PJ}. Consequently, the relations for velocities and accelerations are simply $\dot{\bm{q}}_I=\bm{T}_I\dot{\bm{q}}$ and $\ddot{\bm{q}}_I=\bm{T}_I\ddot{\bm{q}}$, respectively. On the other hand, the variation should exclude the prescribed coordinates as
\begin{equation}\label{eq:varI}
	\delta \bm{q}_I=\check{\bm{T}}_I\delta \check{\bm{q}}.
\end{equation}

Note that the relations \cref{eq:separate,eq:select} are actually implemented as index-selecting methods in the computer code so that expensive matrix multiplications are avoided.

Last but not the least, 
any intrinsic constrains in \cref{eq:bar_intrinsic,eq:3D_intrinsic} and 
extrinsic constrains in \cref{eq:ex_con} that contain no free coordinates should be dropped. 
The remaining constraints are collected by $\check{\bm{\varPhi}}(\bm{q})$, 
whose Jacobian matrix is defined by $\check{\bm{A}}(\bm{q})=\partial\check{\bm{\varPhi}}/{\partial\check{\bm{q}}}$.

\subsection{Generalized forces}
\begin{figure}[tbh]
	\centering
	\includegraphics[width=163pt]{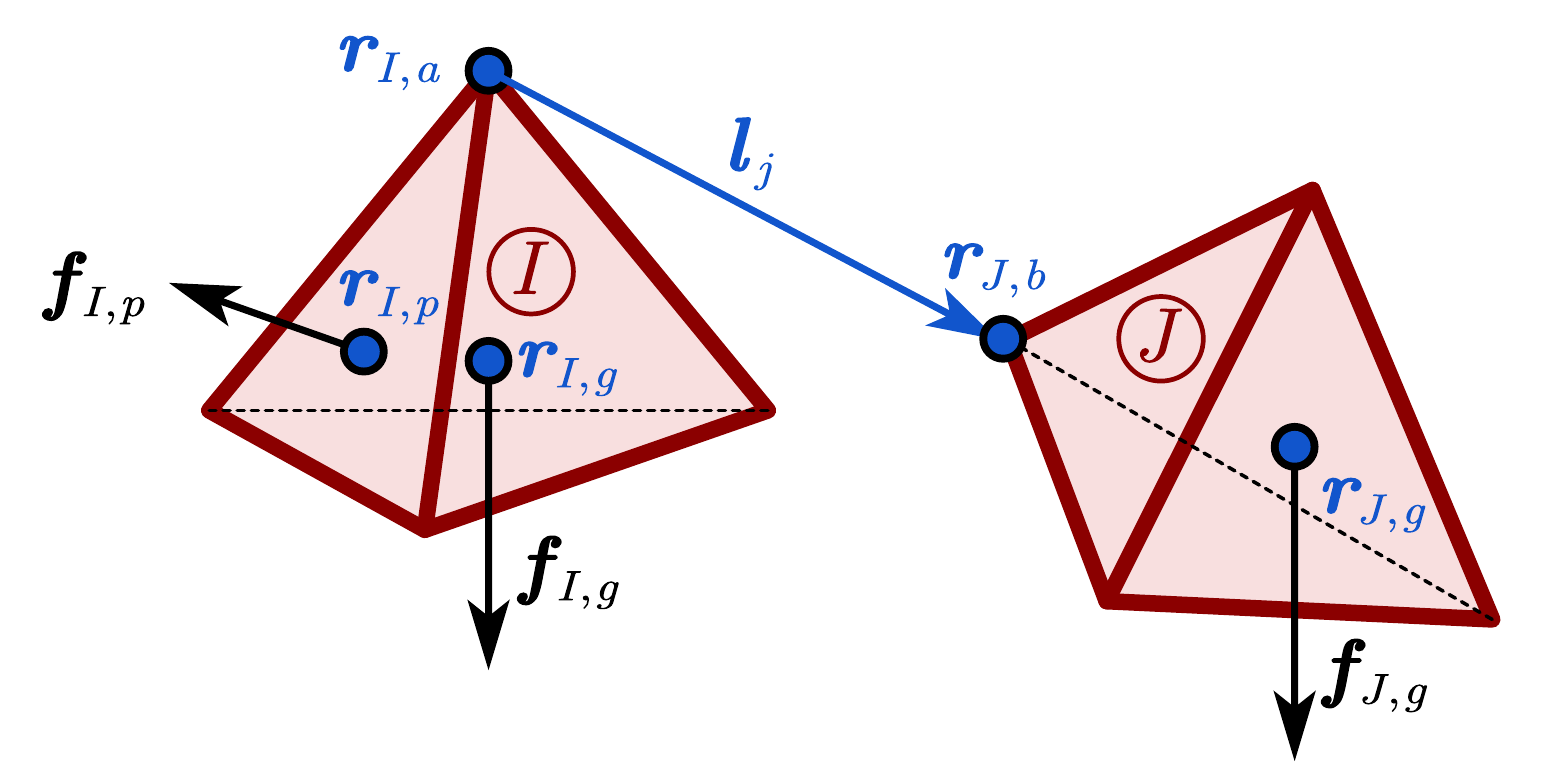}
	\caption{Two 3D rigid bodies subjected to gravity, a concentrated force, and tension forces of a cable. The points of action are colored in blue.}
	\label{fig:GFTF}
\end{figure}
Using \cref{eq:trans,eq:select,eq:varI}, the position and its variation of a point of action $p$ on the rigid member $\CircledSmall{I}$ are, respectively, 
\begin{equation}\label{eq:point}
	\bm{r}_{I,p}=\bm{C}_{I,p}\bm{T}_I\bm{q}\ \ \mathrm{and}\ \ 	\delta\bm{r}_{I,p}=\bm{C}_{I,p}\check{\bm{T}}_I\delta\check{\bm{q}}.
\end{equation}
Consider a concentrated force $\bm{f}_{I,p}$ exerted on point $p$, as shown on the left of \cref{fig:GFTF}, the virtual work done by $\bm{f}_{I,p}$ is $\delta W_{I,p}=\delta \bm{r}_{I,p}^\mathrm{T}\bm{f}_{I,p}=\delta\check{\bm{q}}^\mathrm{T}\check{\bm{F}}_{I,p}$, where
\begin{equation}\label{eq:gen_force}
	\check{\bm{F}}_{I,p}=\check{\bm{T}}^\mathrm{T}_I\bm{C}_{I,p}^\mathrm{T}\bm{f}_{I,p}
\end{equation}
is the generalized force for $\bm{f}_{I,p}$. 

In particular, the gravity force $\bm{f}_{I,g}$ is exerted on the mass center $\bm{r}_{I,g}$. 
Therefore, the generalized gravity force for the rigid member $\CircledSmall{I}$ 
is given by $\check{\bm{F}}_{I,g}=\check{\bm{T}}^\mathrm{T}_I\bm{C}_{I,g}^\mathrm{T}\bm{f}_{I,g}$, 
which is a constant vector. 
\subsection{Tensile cables}
In this paper, we adopt a common practice 
\cite{cefaloComprehensiveDynamicModel2011,cheongNonminimalDynamicsGeneral2015,bohmApproachDynamicsControl2015} 
which assumes that the cables are massless, so that their inertia forces are ignored and 
only their tension forces acting on the rigid members require formulations.

Suppose the $j$th cable connects point $a$ of the rigid member $\CircledSmall{I}$ 
and point $b$ of the rigid member $\CircledSmall{J}$, as shown in \cref{fig:GFTF}. It can be represented by a vector 
\begin{equation}\label{key}
	\bm{l}_j=\bm{r}_{J,b}-\bm{r}_{I,a}=\bm{C}_{J,b}\bm{T}_J\bm{q}-\bm{C}_{I,a}\bm{T}_I\bm{q}=\bm{J}_j\bm{q}
\end{equation}
where we use \cref{eq:point} and $\bm{J}_j=\bm{C}_{J,b}\bm{T}_J-\bm{C}_{I,a}\bm{T}_I$ is a constant matrix. Consequently, the current length and its time derivative of the cable are given by, respectively,
\begin{equation}\label{key}
	l_j=\sqrt{\bm{l}_j^\mathrm{T}\bm{l}_j}
	=\sqrt{\bm{q}^\mathrm{T}\bm{U}_j\bm{q}}\ \ \mathrm{and}\ \ 
	\dot{l}_j
	=\frac{\bm{l}_j^\mathrm{T}\bm{\dot{l}}_j}{\sqrt{\smash[b]{\bm{l}_j^\mathrm{T}\bm{l}_j}}}
	=\frac{\left(\bm{q}^\mathrm{T}\bm{U}_j\bm{\dot{q}}\right)}{l_j},
\end{equation}
where $\bm{U}_j=\bm{J}_j^\mathrm{T}\bm{J}_j$ is also constant.

Define the force density by $\gamma _j={f_j/{l_j}}$, where $f_j$ is the tension force magnitude. Then, the tension force is given by either $\bm{f}_j=f_j\hat{\bm{l}}_j$ or  $\bm{f}_j=\gamma_j\bm{l}_j$, where $\hat{\bm{l}}_j=\bm{l}_j/{l_j}$ is the unit direction vector. 

Note that a cable generates a pair of tension forces exerted on points $a$ and $b$ with opposite directions. Therefore, according to \cref{eq:gen_force}, the generalized tension force for the $j$th cable reads
\begin{equation}\label{key}
	\check{\bm{Q}}_j=
	\check{\bm{T}}^\mathrm{T}_I\bm{C}_{I,a}^\mathrm{T}\bm{f}_j-
	\check{\bm{T}}^\mathrm{T}_J\bm{C}_{J,b}^\mathrm{T}\bm{f}_j
	=-\check{\bm{E}}^\mathrm{T}\bm{J}_j^\mathrm{T}\bm{f}_j
\end{equation}
Consequently, the system's generalized tension force is the sum over all cables
\begin{equation}\label{eq:tension_force}
	\check{\bm{Q}}
	=\sum\nolimits_{j=1}^{n_s}{\left(-\check{\bm{E}}^\mathrm{T}\bm{J}_j^\mathrm{T}\bm{f}_j\right)}
	=-\check{\bm{E}}^\mathrm{T}\mathop{\oplus}_{j=1}^{n_s}(\bm{J}^\mathrm{T}_j\bm{l}_j)\bm{\gamma}
\end{equation}
where $\bm{\gamma}=[\gamma_1,\cdots,\gamma_{n_s}]^\mathrm{T}$ collects the force densities and $\oplus$ means the direct sum of matrices. Expression \cref{eq:tension_force} shows the system's generalized tension force is linear in the cables' force densities. This notable property is also found in the dynamics framework for \enquote{bars-only} tensegrities by Skelton et al. \cite{nagaseNetworkVectorForms2014,cheongNonminimalDynamicsGeneral2015}. It is beneficial for the design of cable-based control schemes, which, however, will not be elaborated in this paper and subject to further research.

Expression \cref{eq:tension_force} allows any constitutive laws of the cables. 
Following common practices, we assume linear stiffness, linear damping, and a slacking behavior. Denote the rest length by $\mu_j$, the stiffness coefficient by $\kappa_j$, and the damping coefficient by $\eta_j$. Then, the tension force magnitude is given by
\begin{equation}\label{eq:constitutive}
	f_j=\left\{ \begin{matrix*}[l]
		f^+_j,&\mathrm{if}\ f^+_j\ge 0\ \ \mathrm{and}\ \ l_j\ge\mu_j\\	
		0,&\mathrm{else}\\
	\end{matrix*} \right.\ \ \mathrm{with}\ \ f^+_j=\kappa_j( l_j-\mu_j ) + \eta_j\dot{l}_j.
\end{equation}

\section{Dynamic analysis formulations}\label{sec:dynamic}

\subsection{Dynamic equation}
Recalling the rigid member's kinetic energy \cref{eq:T_I} and the coordinate selection \cref{eq:select}, the system's kinetic energy is simply the sum over all rigid member $T=\sum_{I=1}^{n_b}T_I=\frac{1}{2}\dot{\bm{q}}^\mathrm{T}\bm{M}\dot{\bm{q}}$, where $\bm{M}=\sum_{I=1}^{n_b}\bm{T}_I^{\mathrm{T}}\bm{M}_I\bm{T}_I$ is constant mass matrix. 
Then, the generalized inertial force is derived with respect to the free coordinates:
\begin{equation}\label{eq:inertia_force}
	\frac{\mathrm{d}}{\mathrm{d}t}\left( \frac{\partial T}{\partial \dot{\check{\bm{q}}}^{\mathrm{T}}} \right) 
	=\frac{\mathrm{d}}{\mathrm{d}t}\left( 
	\acute{\bm{M}}\dot{\bm{q}}
	\right)
	=\frac{\mathrm{d}}{\mathrm{d}t}\left( 
	\acute{\bm{M}}\left( \check{\bm{E}}\dot{\check{\bm{q}}}+\tilde{\bm{E}}\dot{\tilde{\bm{q}}} \right) 
	\right)
	=\check{\bm{M}}\ddot{\check{\bm{q}}}+\bar{\bm{M}}\ddot{\tilde{\bm{q}}}
\end{equation}
where $\acute{\bm{M}}=\check{\bm{E}}^{\mathrm{T}}\bm{M}$, $\check{\bm{M}}=\acute{\bm{M}}\check{\bm{E}}$, and $\bar{\bm{M}}=\acute{\bm{M}}\tilde{\bm{E}}$ are different mass matrices that will be used later.

Suppose a potential $V(\bm{q}$) is given as a function of the system's total coordinates, 
then the generalized potential force in free coordinates is given by $\check{\bm{G}} =	-{\partial V( \bm{q} )}/{\partial \check{\bm{q}}^{\mathrm{T}}}$. 
Furthermore, define $\check{\bm{F}}=\check{\bm{G}}+\check{\bm{Q}}+\check{\bm{F}}^\mathrm{ex}$ to include the generalized potential force $\check{\bm{G}}$, the generalized tension force $\check{\bm{Q}}$, and any other external generalized forces $\check{\bm{F}}^\mathrm{ex}$.

For the dynamics of a tensegrity structure, the Lagrange-d'Alembert principle \cite{greinerClassicalMechanics2010} states that the virtual work vanishes for all inertial forces, generalized forces, and constraint forces acting on the virtual displacement:
\begin{equation}\label{eq:Lagrange}
	\delta \check{\bm{q}}^\mathrm{T}\left( \check{\bm{M}}\ddot{\check{\bm{q}}}+\bar{\bm{M}}\ddot{\tilde{\bm{q}}} \right) -\delta \check{\bm{q}}^\mathrm{T}\check{\bm{F}} - \delta \check{\bm{q}}^\mathrm{T}\left(\check{\bm{A}}^{\mathrm{T}}\bm{\lambda }\right)=0
\end{equation}
which leads to the Lagrange's equation of the first kind 
\begin{subequations}\label{eq:dynamic}
	\begin{empheq}[left=\empheqlbrace]{align}
		\check{\bm{M}}\ddot{\check{\bm{q}}} + \bar{\bm{M}}\ddot{\tilde{\bm{q}}} -
		\check{\bm{G}}(\bm{q})-
		\check{\bm{Q}}(\bm{q},\bm{\dot{q}},\bm{\mu})-
		\check{\bm{F}}^\mathrm{ex}(\bm{q},\dot{\bm{q}},t) - 
		\check{\bm{A}}^{\mathrm{T}}(\bm{q})\bm{\lambda }=\bm{0}\label{eq:dynamic_diff}\\	
		\check{\bm{\varPhi}}(\bm{q})=\bm{0}\label{eq:dynamic_alg}
	\end{empheq}
\end{subequations}
where the dependency is explicated, and the rest lengths $\bm{\mu}$ 
will be used as cable-based actuation values.
One should also keep in mind that $\bm{q}$ contains prescribed coordinates $\tilde{\bm{q}}$, which, along with $\dot{\tilde{\bm{q}}}$ and $\ddot{\tilde{\bm{q}}}$, are interpreted as known functions of time $t$.

Thanks to the use of natural coordinates, the dynamic equation \cref{eq:dynamic} gets rid of trigonometric functions as well as inertia quadratic velocity terms for centrifugal and Coriolis forces, leaving a constant mass matrix. 

For later use, the differential part \cref{eq:dynamic_diff} can be rewritten as 
\begin{equation}\label{eq:dynamic_rw}
	\dot{\check{\bm{p}}}-\check{\bm{F}}-\check{\bm{A}}^{\mathrm{T}}\bm{\lambda}=\bm{0}
\end{equation}
where $\check{\bm{p}}={\partial T}/{\partial\dot{\check{\bm{q}}}^{\mathrm{T}}}=\acute{\bm{M}}\dot{\bm{q}}$ 
is the generalized momentum in free coordinates.

\subsection{Linearized dynamics around static equilibrium}\label{sec:linearized}
In order to perform modal analysis on general tensegrity structures,
this subsection derives the formulations of linearized dynamics around static equilibrium.

Dropping all time-related terms in the dynamic equation \cref{eq:dynamic} leads to the static equation
\begin{subequations}\label{eq:static}
	\begin{empheq}[left=\empheqlbrace]{align}
		-\check{\bm{F}}(\bm{q}) 
		-\check{\bm{A}}^{\mathrm{T}}(\bm{q})\bm{\lambda }
		=\bm{0}\label{eq:static_diff}\\	
		\check{\bm{\varPhi}}(\bm{q})
		=\bm{0}\label{eq:static_alg}
	\end{empheq}
\end{subequations}
Consider small perturbations in the free coordinates and Lagrange multipliers as
\begin{equation}\label{eq:perturbation}
	\bm{q}=\bm{q}_\mathrm{e}+\check{\bm{E}}\delta\check{\bm{q}},\ \ 
	\dot{\bm{q}}=\dot{\bm{q}}_\mathrm{e}+\check{\bm{E}}\delta\dot{\check{\bm{q}}},\ \ 
	\ddot{\bm{q}}=\ddot{\bm{q}}_\mathrm{e}+\check{\bm{E}}\delta\ddot{\check{\bm{q}}},\ \ \mathrm{and}\  \
	\bm{\lambda}=\bm{\lambda}_\mathrm{e}+\delta\bm{\lambda},
\end{equation}
where $\dot{\bm{q}}_\mathrm{e}=\ddot{\bm{q}}_\mathrm{e}=\bm{0}$, and $(\bm{q}_\mathrm{e},\bm{\lambda}_\mathrm{e})$ satisfies the static equation \cref{eq:static}. Substituting \cref{eq:perturbation} into \cref{eq:dynamic} and expanding it in Taylor series to the first order lead to
\begin{subequations}\label{eq:linearized}
	\begin{empheq}[left=\empheqlbrace]{align}
		\check{\bm{M}}\delta \ddot{\check{\bm{q}}}-
		\frac{\partial \check{\bm{F}}}{\partial \dot{\check{\bm{q}}}}\delta \dot{\check{\bm{q}}}-
		\frac{\partial \check{\bm{F}}}{\partial \check{\bm{q}}}\delta \check{\bm{q}}-
		\frac{\partial \left( \check{\bm{A}}^\mathrm{T}\bm{\lambda }_\mathrm{e} \right)}{\partial \check{\bm{q}}}\delta \check{\bm{q}}-
		\check{\bm{A}}^\mathrm{T}\delta \bm{\lambda }&=\bm{0}\label{eq:linearized_dif}\\[-5pt]
		\check{\bm{A}}\delta \check{\bm{q}}&=\bm{0}\label{eq:linearized_alg}
	\end{empheq} 
\end{subequations}
Define $\check{\bm{N}}$ as a basis of the nullspace 
$\mathcal{N}(\check{\bm A})=\{\bm x|\check{\bm A}\bm x=\bm 0\}$.
So \cref{eq:linearized_alg} is solved by 
\begin{equation}
	\delta\check{\bm{q}}=\check{\bm{N}}\bm{\xi},
	\label{eq:deltaq}
\end{equation}
where $\bm{\xi}\in \mathbb{R}^{n_{\mathrm{dof}}}$ are independent variations 
and $ n_{\mathrm{dof}}$ denotes the degrees of freedom.
Left-multiplying \cref{eq:linearized_dif} by $\check{\bm{N}}^\mathrm{T}$ and 
substituting \cref{eq:deltaq} to \cref{eq:linearized_dif} gives
\begin{equation}\label{eq:reduced_basis_linearized}
	\bm{\mathcal{M}}\ddot{\bm{\xi}}+
	\bm{\mathcal{C}}\dot{\bm{\xi}}+
	\bm{\mathcal{K}}\bm{\xi}
	=\bm{0}
\end{equation}
where
\begin{equation}\label{key}	
	\bm{\mathcal{M}} = \check{\bm{N}}^\mathrm{T}\check{\bm{M}}\check{\bm{N}},\ 
	\bm{\mathcal{C}} = \check{\bm{N}}^\mathrm{T}\left( \frac{-\partial\check{\bm{F}}}{\partial \dot{\check{\bm{q}}}}\right)\check{\bm{N}},\ \mathrm{and} \ \ 
	\bm{\mathcal{K}} = \check{\bm{N}}^\mathrm{T}\left(\frac{-\partial \check{\bm{F}}}{\partial \check{\bm{q}}}-
	\frac{\partial \left( \check{\bm{A}}^\mathrm{T}\bm{\lambda }_\mathrm{e} \right)}{\partial \check{\bm{q}}}\right)\check{\bm{N}}
\end{equation}
are the reduced-basis mass matrix, reduced-basis tangent damping matrix, and reduced-basis tangent stiffness matrix, respectively. 
Such operations are known as the reduced basis method 
\cite{erikssonConstrainedStabilityConservative2019} 
and the nullspace matrix $\check{\bm{N}}$ can be computed by the singular value decomposition of $\check{\bm{A}}$.

At this point, we have a standard linear dynamic system \cref{eq:reduced_basis_linearized}, 
which can be used for the modal analysis of general tensegrity structures. 
For simplicity, consider undamped free vibration ($\bm{\mathcal{C}}=\bm{0}$), 
then the solution to \cref{eq:reduced_basis_linearized} boils down to the generalized eigenvalue problem  
\begin{equation}\label{eq:eigen}
	\left(\bm{\mathcal{K}}-\rho_{(r)}\bm{\mathcal{M}}\right)\bm{\xi}_{(r)}=\bm{0}
\end{equation}
where $ \rho_{(r)} $ is the $r$th eigenvalue in the order of increasing magnitude, 
and $ \bm{\xi}_{(r)} $ is the corresponding eigenvector.
According to the Lyapunov theorem on stability in the first approximation,
the structure's stability around static equilibrium is guaranteed by the positiveness of the lowest eigenvalue:
\begin{equation}\label{eq:stability}
	\rho_{(1)} > 0
\end{equation}

For a detailed exposition of the static stability of constrained structures, we refer the interested readers to Ref. \cite{erikssonConstrainedStabilityConservative2019}. 
Once the stability criterion \cref{eq:stability} is met, 
we can calculate the natural frequency of the $r$th mode by $\omega_{(r)} = \sqrt{\rho_{(r)}}$,
and normalize the mode shape with respect to mass by $\hat{\bm{\xi}}_{(r)} = \frac{1}{\sqrt{m_{(r)}}}\bm{\xi}_{(r)}$, 
where $m_{(r)}=\bm{\xi}^\mathrm{T}_{(r)}\bm{\mathcal{M}}\bm{\xi}_{(r)}$. 
Then, the mode shapes in the natural coordinates can be obtained through
\begin{equation}\label{eq:mode_shapes}
	\bm{q}_{(r)} = \bm{q}_e + \check{\bm{E}}\delta\check{\bm{q}}_{(r)} = \bm{q}_e + \check{\bm{E}}\check{\bm{N}}\hat{\bm{\xi}}_{(r)}
\end{equation}

\subsection{Modified symplectic integration scheme for nonlinear dynamics}
Consider deployable tensegrity structures, 
such as tensegrity space booms \cite{yildizSizingPrestressOptimization2020} and 
tensegrity footbridge \cite{belhadjaliDesignOptimizationDynamic2010},
which are capable to achieve large-range movements under cable-based actuation.
The deployment process would take a sufficiently long time for safety reasons,
but still exhibits rich behaviors \cite{kanNonlinearDynamicDeployment2018a} 
due to the complex rigid-tensile coupling in tensegrity dynamics.
Therefore, when developing solution methods for the governing DAEs \cref{eq:dynamic}, 
attentions should be paid to the numerical performances in long-time simulations.
In this regard, we adopt the Zu-class symplectic integration method
\cite{zhongIntegrationConstrainedDynamical2006,wuConstrainedHamiltonVariational2016} 
which have advantages in two aspects:
Firstly, it can produce realistic results with relatively large timesteps, 
because it preserves the symplectic map of conservative systems; 
it has no artificial dissipation; and it enforces the algebraic constraints;
Secondly, it dispenses with the computations of accelerations 
(and acceleration-like variables as in the generalized-$\alpha$ method 
\cite{arnoldConvergenceGeneralizedaScheme2007}) 
and the partial derivatives of the constraint force. 
Hence, the Zu-class method excels in numerical accuracy and efficiency for long-time simulations.
Nonetheless, it did not originally accommodate non-conservative forces and 
boundary conditions that are present in the governing DAEs \cref{eq:dynamic}.
To address these issues, a rework from the viewpoint of approximations and limits 
are carried out as follows. 

\begin{figure}[tbh]
	\centering
	\includegraphics[width=206pt]{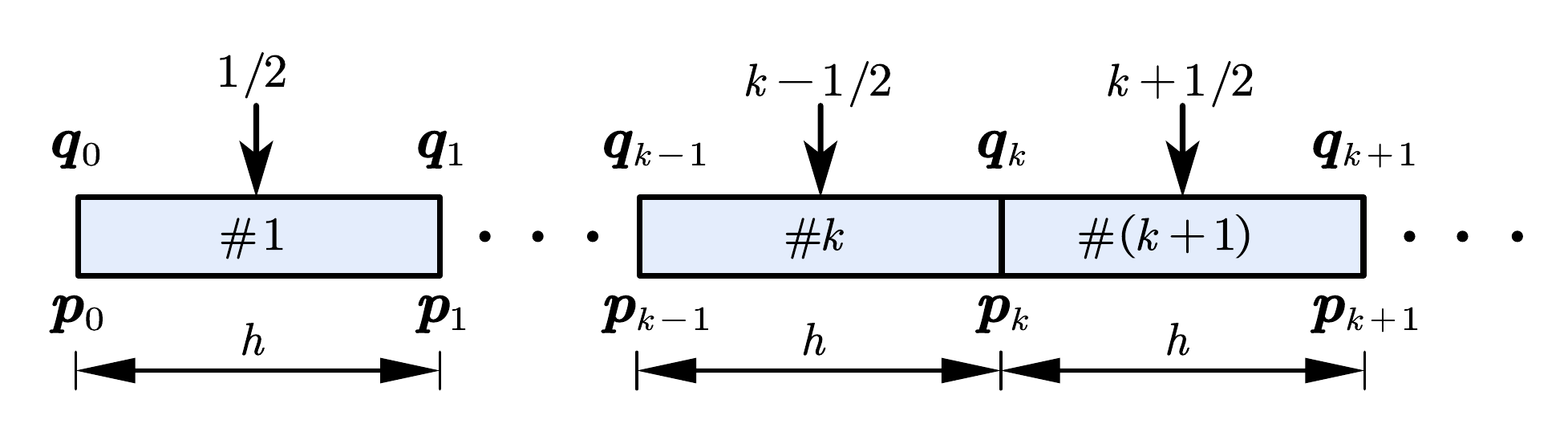}
	\caption{Equally spaced segments of the time domain. Each segment has two endpoints and one midpoint. The state vector $\left( \bm{q}_k,\bm{p}_k \right)$ is located at endpoint $k$.}
	\label{fig:scheme}
\end{figure}

As illustrated in \cref{fig:scheme}, the time domain is divided into equally spaced segments, where $h$ is the timestep and $\left( \bm{q}_k,\bm{p}_k \right) $ denotes the state vector at the segments' endpoints. At each endpoint, we demand that the differential equation \cref{eq:dynamic_rw} holds as 
\begin{equation}\label{eq:dynamic_rw_k}
	\dot{\check{\bm{p}}}_k-\check{\bm{F}}_k-\check{\bm{A}}^{\mathrm{T}}(\bm{q}_k)\bm{\lambda}_k=\bm{0}
\end{equation}
Then, substituting central difference approximations 
\begin{equation}\label{eq:cd_approx}
	\begin{aligned}
		\dot{\check{\bm{p}}}_k&\approx \tfrac{1}{h}\left(\check{\bm{p}}_{k+1/2}-\check{\bm{p}}_{k}+\check{\bm{p}}_{k}-\check{\bm{p}}_{k-1/2}\right), \ \ \check{\bm{F}}_k\approx \tfrac{1}{2}\left(\check{\bm{F}}_{k-1/2}+\check{\bm{F}}_{k+1/2}\right),\ \mathrm{and}\\
		\bm{\lambda}_k&\approx \tfrac{1}{2}\left(\bm{\lambda}_{k-1/2}+\bm{\lambda}_{k+1/2}\right)
	\end{aligned}
\end{equation}
into \cref{eq:dynamic_rw_k} leads to a discrete scheme
\begin{equation}\label{eq:two_step_scheme}
	\frac{\check{\bm{p}}_{k+1/2}-\check{\bm{p}}_{k}+\check{\bm{p}}_{k}-\check{\bm{p}}_{k-1/2}}{h}-
	\frac{\check{\bm{F}}_{k-1/2}+\check{\bm{F}}_{k+1/2}}{2}-
	\check{\bm{A}}(\bm{q}_k)^\mathrm{T}\tfrac{\bm{\lambda}_{k-1/2}+\bm{\lambda}_{k+1/2}}{2}=\bm{0}
\end{equation}
where the midpoint approximations are 
\begin{equation}\label{eq:midpoint_approx}
	\begin{aligned}
		\bm{q}_{k+1/2}&\approx \tfrac{1}{2}(\bm{q}_k+\bm{q}_{k+1}),\ \ 
		\dot{\bm{q}}_{k+1/2}\approx \tfrac{1}{h}(\bm{q}_{k+1}-\bm{q}_k),\\
		\check{\bm{p}}_{k+1/2}&\approx\tfrac{1}{h}\acute{\bm{M}}(\bm{q}_{k+1}-\bm{q}_k),\ \ \mathrm{and}\ \ 
		\check{\bm{F}}_{k+1/2}\approx\check{\bm{F}}\left( \bm{q}_{k+1/2},\dot{\bm{q}}_{k+1/2},t_{k+1/2} \right) \\
	\end{aligned}
\end{equation}

Note that \cref{eq:two_step_scheme} is actually a two-timestep scheme, 
but can be converted to a one-timestep scheme. 
As illustrated in \cref{fig:scheme}, 
the scheme \cref{eq:two_step_scheme} at endpoint $k$ have terms in both segments 
$\#k$ and $\#(k{+}1)$. Taking the limit $t_{k-1}\to t_k$, we have 
\begin{equation}\label{key}
	\lim_{h\to 0} \tfrac{\check{\bm{p}}_k-\check{\bm{p}}_{k-1/2}}{h/2}=\dot{\check{\bm{p}}}_k,\ \ \lim_{h\to 0} \check{\bm{F}}_{k-1/2}=\check{\bm{F}}_k,\ \ \mathrm{and}\ \ 
	\lim_{h\to 0} \bm{\lambda}_{k-1/2}=\bm{\lambda}_k
\end{equation}
which shows that the terms in segment $\#k$ tend to \cref{eq:dynamic_rw_k}, so they can be dropped, leaving
\begin{equation}\label{eq:endpoint_right}
	\check{\bm{p}}_{k+1/2}-\check{\bm{p}}_k-\tfrac{h}{2}\check{\bm{F}}_{k+1/2}-{ \tfrac{h}{2}\check{\bm{A}}(\bm{q}_k)^\mathrm{T}\bm{\lambda }_{k+1/2}}=\bm{0}
\end{equation}
Similarly, taking the limit $t_{k+1}\to t_k$ in \cref{eq:two_step_scheme} leads to 
\begin{equation}\label{eq:endpoint_left}	
	\check{\bm{p}}_k-\check{\bm{p}}_{k-1/2}-\tfrac{h}{2}\check{\bm{F}}_{k-1/2}-{ \tfrac{h}{2}\check{\bm{A}}(\bm{q}_k)^\mathrm{T}\bm{\lambda }_{k-1/2}}=\bm{0}
\end{equation}
Then, applying \cref{eq:endpoint_left} to endpoint $k{+}1$, and combining it with \cref{eq:endpoint_right} as well as the constraint equations, lead to a new scheme:
\begin{subequations}\label{eq:new_scheme}
	\begin{empheq}[left=\empheqlbrace]{align}
		\tfrac{1}{h}{\acute{\bm{M}}( \bm{q}_{k+1}{-}\bm{q}_{k} )}-\check{\bm{p}}_k-\tfrac{h}{2}\check{\bm{F}}_{k+1/2}-{ \tfrac{h}{2}\check{\bm{A}}(\bm{q}_k)^\mathrm{T}\bm{\lambda }_{k+1/2}}&=\bm{0}\label{eq:new_scheme_k}\\
		\check{\bm{p}}_{k+1}-\tfrac{1}{h}{\acute{\bm{M}}( \bm{q}_{k+1}{-}\bm{q}_{k} )}-\tfrac{h}{2}\check{\bm{F}}_{k+1/2}-{ \tfrac{h}{2}\check{\bm{A}}(\bm{q}_{k+1})^\mathrm{T}\bm{\lambda }_{k+1/2}}&=\bm{0}\label{eq:new_scheme_kp1}\\
		\check{\bm{\varPhi}}( \bm{q}_{k+1} )&=\bm{0}\label{eq:new_scheme_alg}
	\end{empheq}
\end{subequations}
We call it the modified symplectic integration (MSI) scheme, because it automatically includes boundary conditions through prescribed coordinates and allows for non-conservative forces given by $\check{\bm{F}}$. These two aspects were not considered in its original derivations \cite{zhongIntegrationConstrainedDynamical2006}.
To provide a solution procedure, 
rearrange \cref{eq:new_scheme_k,eq:new_scheme_alg} as a residual expression
\begin{equation}\label{eq:Res}
	\mathbf{Res}(\bm{x}_{k+1})=
	\begin{pmatrix}\textstyle
		-h\check{\bm{p}}_{k}+
		\acute{\bm{M}}\left( \bm{q}_{k+1}-\bm{q}_{k} \right)-\frac{h^2}{2}\check{\bm{F}}_{k+1/2}-s_1\frac{h^2}{2}\check{\bm{A}}^\mathrm{T}( \bm{q}_{k} ) \bm{\lambda }_{k+{1}/{2}}\\	
		\check{\bm{\varPhi}}( \bm{q}_{k+1} )\\
	\end{pmatrix}
\end{equation}
where $\bm{x}_{k+1}{=}[\check{\bm{q}}_{k+1}^\mathrm{T},\bm{\lambda }_{k+1/2}^\mathrm{T}] ^\mathrm{T}$, and $s_1{=}2h^{-2}$ is a scaling factor \cite{bauchauScalingConstraintsAugmented2009} that is needed for better conditioning of the Jacobian matrix 
\begin{equation}\label{eq:ResJac}
	\mathbf{Jac}(\bm{x}_{k+1})=\frac{\partial \mathbf{Res}}{\partial \bm{x}_{k+1}}=
	\begin{bmatrix}	
		\check{\bm{M}}-\frac{h^2}{2}\frac{\partial \check{\bm{F}}_{k+1/2}}{\partial \check{\bm{q}}_{k+1}}&		-\check{\bm{A}}^\mathrm{T}( \bm{q}_{k} )\\	\check{\bm{A}}( \bm{q}_{k+1} )&		\bm{0}\\
	\end{bmatrix}
\end{equation}
The residual \cref{eq:Res} and its Jacobian \cref{eq:ResJac} allow us to solve for $\bm{x}_{k+1}$ using the Newton-Raphson iteration method. After that, $\bm{x}_{k+1}$ is substituted into \cref{eq:new_scheme_kp1} to compute $\check{\bm{p}}_{k+1}$ explicitly. 

We can observe that accelerations $\ddot{\bm{q}}$ and partial derivatives of the constraint force $\check{\bm{A}}^{\mathrm{T}}(\bm{q})\bm{\lambda }$, 
which are needed for other schemes \cite{arnoldConvergenceGeneralizedaScheme2007}, 
do not appear in \cref{eq:Res,eq:ResJac}. 


The complete solution procedure of MSI is summarized in \cref{alg:scheme}. 

\begin{algorithm}[tbh!]
\caption{Modified symplectic integration (MSI) scheme}\label{alg:scheme}
\begin{algorithmic}[1]
\algrenewcommand{\algorithmiccomment}[1]{\hskip1em // #1}
\algnewcommand\algorithmicto{\textbf{to}}
\Require {initial values $\bm{q}_0$ and $\dot{\bm{q}}_0$; timestep $h$, total steps $N$, 
maximum iteration $s_\mathrm{max}$, tolerance $\epsilon_\mathrm{tol}$}
\State $\bm{p}_0\gets \bm{M}\dot{\bm{q}}_0$\;
\For{$k\gets 0~\textbf{to}~N-1$}
        \State $\check{\bm{q}}_{k+1} \gets \check{\bm{q}}_{k}$\;
        \State $\bm{\lambda}_{k+1/2} \gets \bm{0}$\;
        \State $\bm{x}_{k+1}\gets [\check{\bm{q}}_{k+1}^\mathrm{T},\bm{\lambda }_{k+1/2}^\mathrm{T}] ^\mathrm{T}$\;
        \For{$s \gets 1~\textbf{to}~s_\mathrm{max}$}\Comment{\small Newton-Raphson iteration}
            \State compute $\mathbf{Res}$ by \cref{eq:Res}\;
            \If{$\lVert\mathbf{Res}\rVert>\epsilon_\mathrm{tol}$}
                \State compute $\mathbf{Jac}$ by \cref{eq:ResJac}\;
                \State $\Delta \bm{x} \gets -(\mathbf{Jac})^{-1}\mathbf{Res}$\;
                \State $\bm{x}_{k+1}\gets \bm{x}_{k+1}+\Delta \bm{x}$\;
            \Else
                \State break\;
            \EndIf
        \EndFor
        \State compute $\check{\bm{p}}_k$ by \cref{eq:new_scheme_kp1}\;
        \State $\dot{\check{\bm{q}}}_k \gets \check{\bm{M}}^{-1}\left(\check{\bm{p}}_k-\tilde{\bm{M}}\dot{\tilde{\bm{q}}}_k\right)$\;
\EndFor
\end{algorithmic}
\end{algorithm}

\section{Numerical examples}\label{sec:examples}

Numerical studies of four representative examples are presented in this section.
The purpose is two-fold: 
(1) To exemplify three-dimensional general tensegrity structures composed of arbitrary rigid bodies and rigid bars; 
(2) To demonstrate the efficacy of the proposed unified approach for dynamic analyses of general tensegrity structures. 

The example structures can be categorized into two groups.
The first group includes examples 1 and 2 that are designed by algorithmic methods, 
such as the topology-finding method \cite{wangTopologyDesignGeneral2020}.
The second group includes examples 3 and 4 that are designed by intuitive methods, 
which will be called the \enquote{embedding} and \enquote{interfacing} methods. 
The connotation of the intuitive methods will be explained in subsections.

The different dynamic behaviors of these structures will be demonstrated,  
and various complex conditions will be considered, including cable-based deployments, and moving boundaries. 
Additionally, the proposed MSI scheme will be compared against the state-of-the-art method.
\subsection{Example 1: A fusiform tensegrity structure}

\begin{figure}[tbh]
	\centering
	\includegraphics[width=0.4\lw]{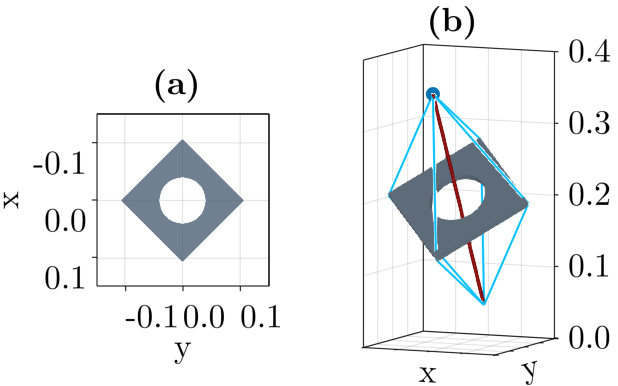}
	\caption{
		(a) Dimensions of a rigid square board; 
		(b) Initial configuration of a fusiform tensegrity structure composed of a rigid bar and a rigid square board.
	}
	\label{fig:simple}
\end{figure}
\begin{figure}[tbh]
	\centering
	\includegraphics[width=0.8\lw]{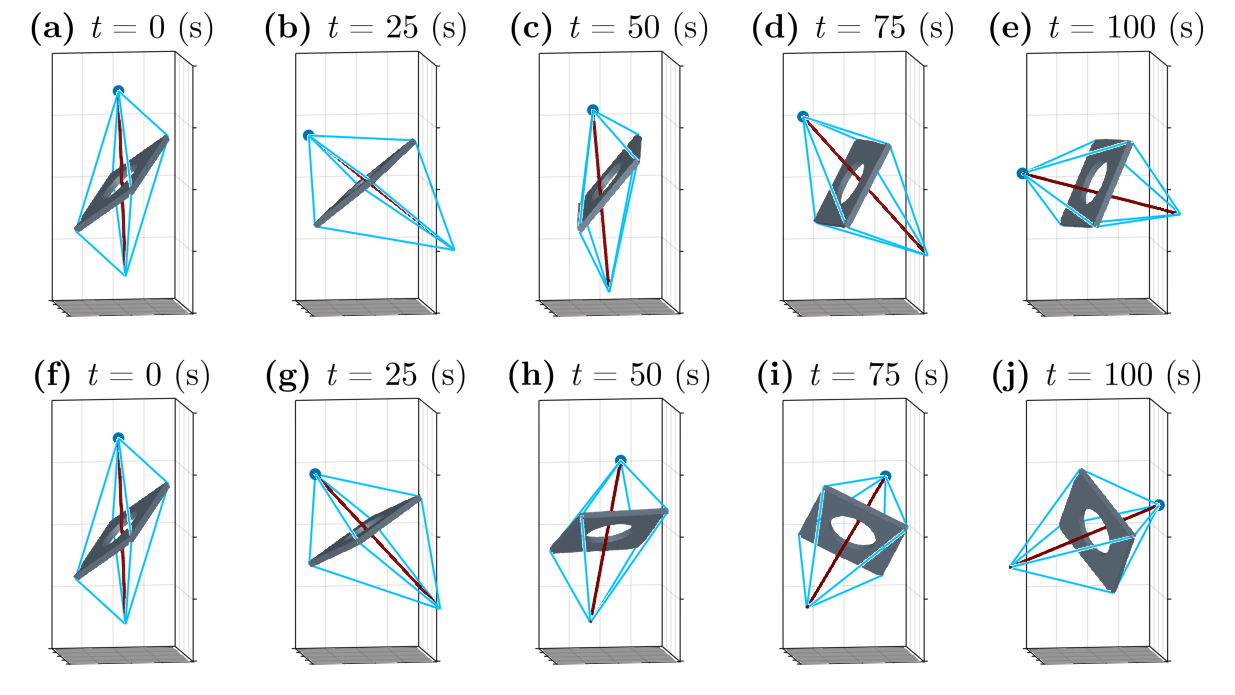}
	\caption{
		Snapshots of the fusiform tensegrity structure at different time instances simulated by 
		(a,b,c,d,e) the MSI scheme and (f,g,h,i,j) the generalized-$\alpha$ scheme.
		Blue dots indicate the marker point. 
	}
	\label{fig:simplesnapshots}
\end{figure}
\begin{figure}[tbh]
	\centering
	\includegraphics[width=0.8\lw]{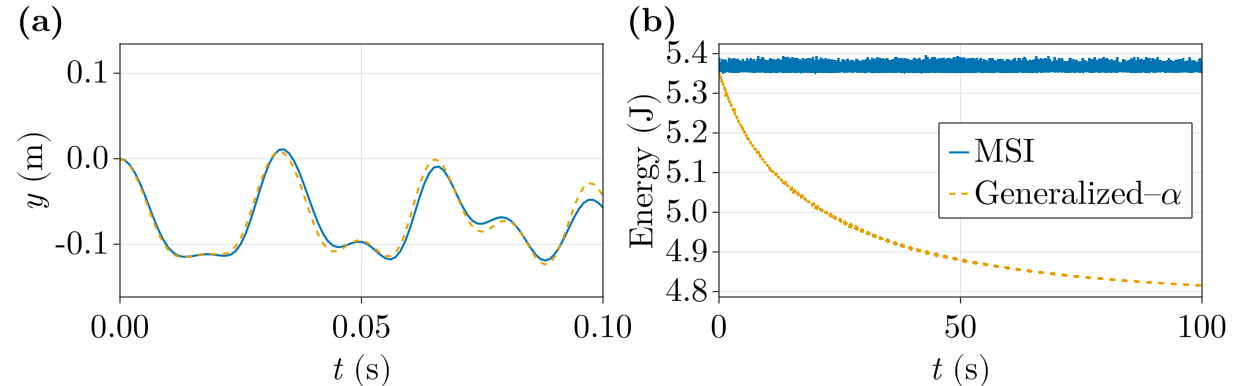}
	\caption{
		(a) Trajectories of the marker point in the $y$-direction and 
		(b) time histories of the mechanical energy $E$ 
		given by the MSI scheme and generalized-$\alpha$ scheme.
	}
	\label{fig:simpleenergy}
\end{figure}
This example considers a three-dimensional fusiform tensegrity structure, 
involving a punctured square rigid board and a rigid bar.
In a study of topology-finding method \cite{wangTopologyDesignGeneral2020}, 
this structure represents one of the simplest Class-1 general tensegrities. 
A variant of this structure, which replaces the punctured square with a triangle,
is studied by Liu et al.~\cite{liuKinematicStaticAnalysis2020} as a tensegrity robot. 
However, due to difficulties arising from the heterogeneity of rigid members, 
the dynamic characteristics of this structure were not studied in the above references.
To demonstrate its rich dynamic motions, an initially unbalanced configuration, 
where the rigid board is rotated around the $x$-axis by $45\degree$, 
and the rigid bar is rotated around the $y$-axis by $15\degree$, 
as shown in \cref{fig:simple} (b).
Both rigid members are given a uniform density $\rho=\SI{630}{\kilogram\per\metre}$, 
corresponding to teak wood.
All eight cables have a stiffness coefficient $\kappa = \SI{100}{\newton\per\metre}$
with no damping.
The upper four cables are given rest length $\mu = \SI{0.05}{\metre} $, 
while the lower four ones have $\mu = \SI{0.1}{\metre}$.
The structure is free-floating.

Consider 100-second long-time simulations with timestep $h = \SI{1e-3}{\second}$,
carried out by the MSI scheme and the generalized-$\alpha$ scheme 
\cite{arnoldConvergenceGeneralizedaScheme2007}.
\cref{fig:simplesnapshots} visualizes the structural movements, 
while \cref{fig:simpleenergy} compares the trajectories of the marker point and 
the mechanical energy $E=T+V$ produced by the two schemes.
These results show that the motions of a 3D rigid bar are correctly described by 
the natural coordinates without any difficulty,
and that the trajectories between the two schemes are very close in the beginning of the simulations.
In particular, the MSI scheme conserves the mechanical energy $E$ 
and obtains vibrations between the two rigid members throughout the entire process.
In contrast, the generalized-$\alpha$ scheme with $\rho_\infty=0.7$ gradually 
damps out such high-frequncy vibrations and dissipates the associated energy.
Therefore, the MSI scheme is more suitable to faithfully simulate the long-time dynamics of 
general tensegrity structures.

\subsection{Example 2: A tensegrity bridge}
\begin{figure}[tbh!]
	\centering
	\includegraphics[width=0.9\lw]{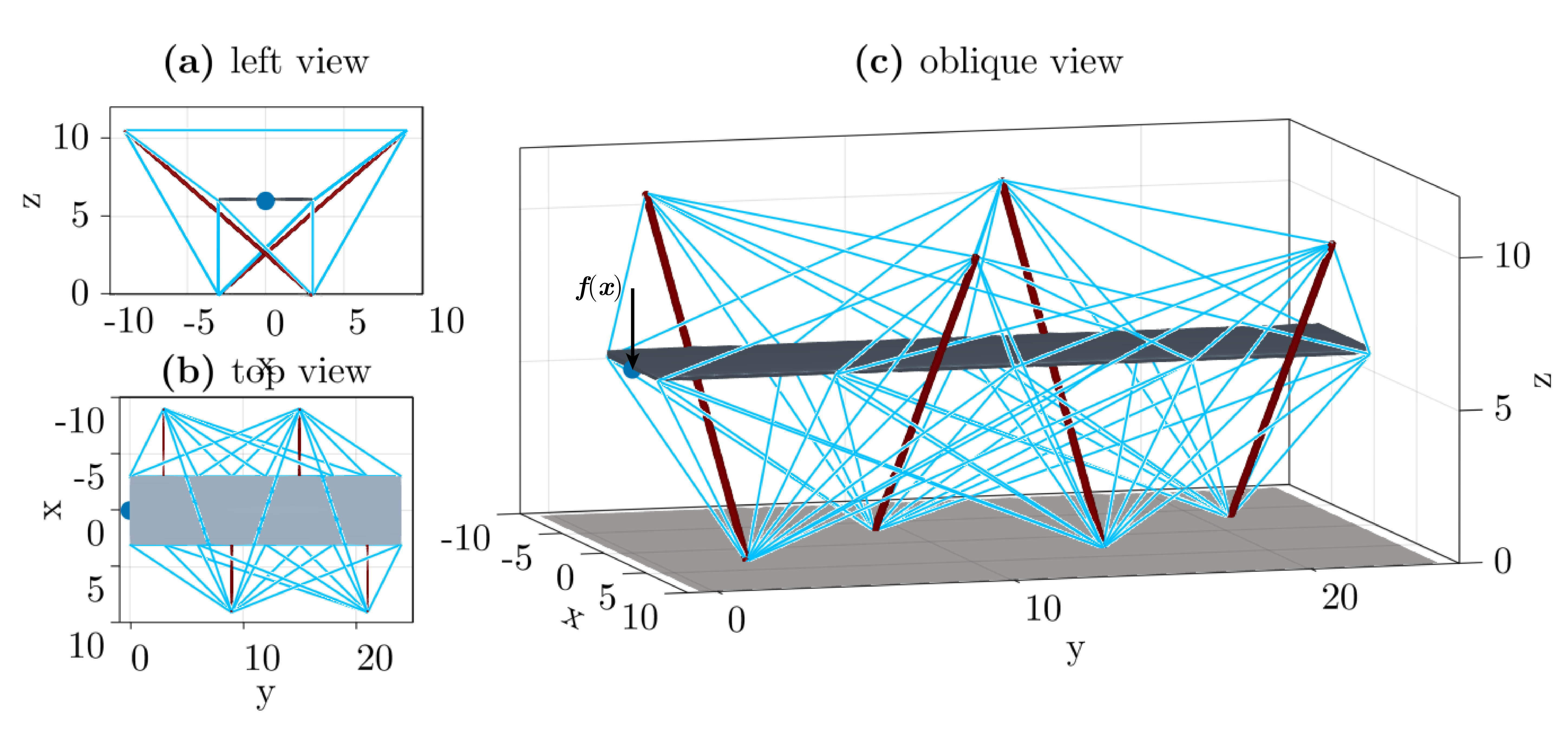}
	\caption{Schematic figures of the tensegrity bridge from 
		(a) left view, 
		(b) top view, 
		and (c) oblique view with a concentrated loading force.
		Blue dots indicate the marker point. 
	}
	\label{fig:bridge}
\end{figure}

\begin{figure}[tbh!]
	\centering
	\includegraphics[width=\lw]{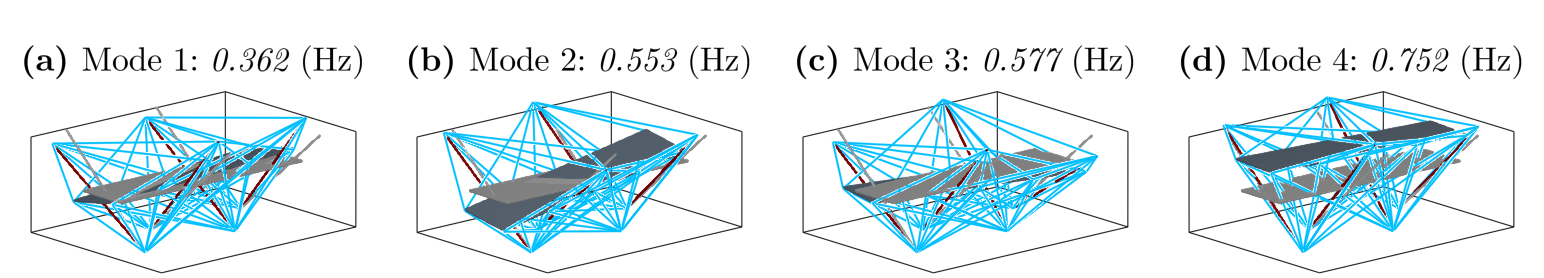}
	\caption{
		The mode shapes and natural frequencies for the first four vibration modes of the tensegrity bridge.
		The configuration of static equilibrium is colored in gray for reference.
	}
	\label{fig:bridgemodes}
\end{figure}

\begin{figure}[tbh!]
	\centering
	\includegraphics[width=0.5\lw]{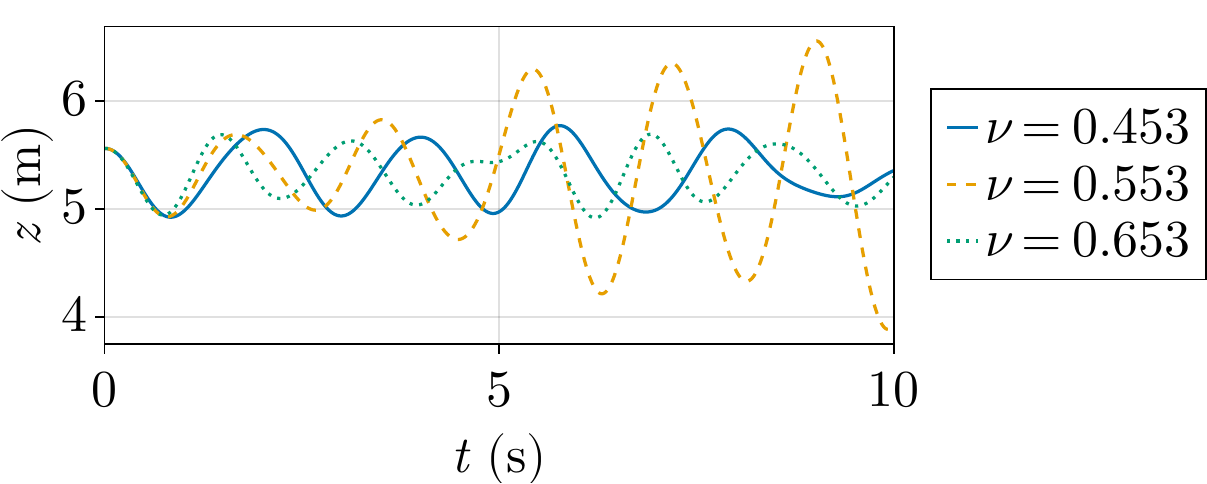}
	\caption{
		Trajectories of the loading point in the $z$-direction for 
		the three simulation cases with different excitation frequencies.
	}
	\label{fig:bridgevibes}
\end{figure}

This example is a Class-1 tensegrity bridge composed of a rectangular rigid body 
as the bridge deck and inclined rigid bars as supporting struts, 
as shown in \cref{fig:bridge}. 
It represents another example resulting from the design method of 
topology-finding \cite{wangTopologyDesignGeneral2020}.
Because the bars have no contact with the deck, it is a class-1 tensegrity structure.
Each rigid bar has a length $l = \SI{15.95}{\meter}$ and a virtual radius of cross-section $r = \SI{0.13}{\meter}$, and 
the deck has dimensions $\SI{24}{\meter}\times \SI{6}{\meter}\times \SI{0.25}{\meter}$ 
(length $\times$ width $\times$ height). 
Note that material properties were not considered in the above reference.
For demonstration purpose, rigid members are given a uniform density of teak wood $\rho = \SI{630}{\kilogram\per\meter^3}$,
and cables are given a stiffness coefficient $\kappa = \SI{25.918}{\kilo\newton\per\meter}$.
Furthermore, the lower end of each bar is fixed to the ground, 
so that the structure can support self-weight and loading forces.

Due to the heterogeneity between rigid bodies and rigid bars, 
it is difficult to obtain reference results for 
the dynamic behaviors of the bridge in commercial software  that uses minimal coordinates,
such as Adams.
Therefore, in order to validate the dynamic formulations and the MSI scheme, 
the resonance phenomenon will be simulated.
Firstly, a static equilibrium configuration and the rest lengths of cables are sought 
by the geodesic dynamic relaxation method \cite{mikiGeodesicDynamicRelaxation2014}. 
Then, linearized dynamic analysis is performed to compute 
the natural frequencies and mode shapes which reveal 
how the structure vibrates around the initial static equilibrium.
The first four vibration modes are shown in \cref{fig:bridgemodes}.
In particular,  a tilting movement of the deck can be observed from the second mode with a natural frequency $\SI{0.553}{\hertz}$.
Based on this observation, the nonlinear dynamics simulations can be validated by inducing vibrations resonating with this frequency.
To this end, a  concentrated loading force $\bm{f}(t)$ with different frequencies is exerted to the edge of the deck, as shown in \cref{fig:bridge} (c).
The force magnitude is a function of time $f(t)= \num{2e4}\left( \sin\left(2\pi \nu t\right) + 1\right) \unit{\newton}$, where $\nu = 0.453, 0.553, 0.653 \unit{\hertz}$ are three excitation frequencies, 
representing three simulation cases. 
Nonlinear dynamic simulations for 10 seconds are performed for each case with timestep $h = \SI{1e-2}{\second}$, using the MSI scheme.
Trajectories of the loading point in the $z$-direction is plotted in \cref{fig:bridgevibes}.
It shows that the amplitude of response is significantly increased only for $\nu = \SI{0.553}{\hertz}$, 
indicating vibrations resonant with the second mode, 
and hence validates the proposed modeling formulations and integration scheme.

\subsection{Example 3: A tensegrity structure designed by embedding}
\begin{figure}[tbh]
	\centering
	\includegraphics[width=\lw]{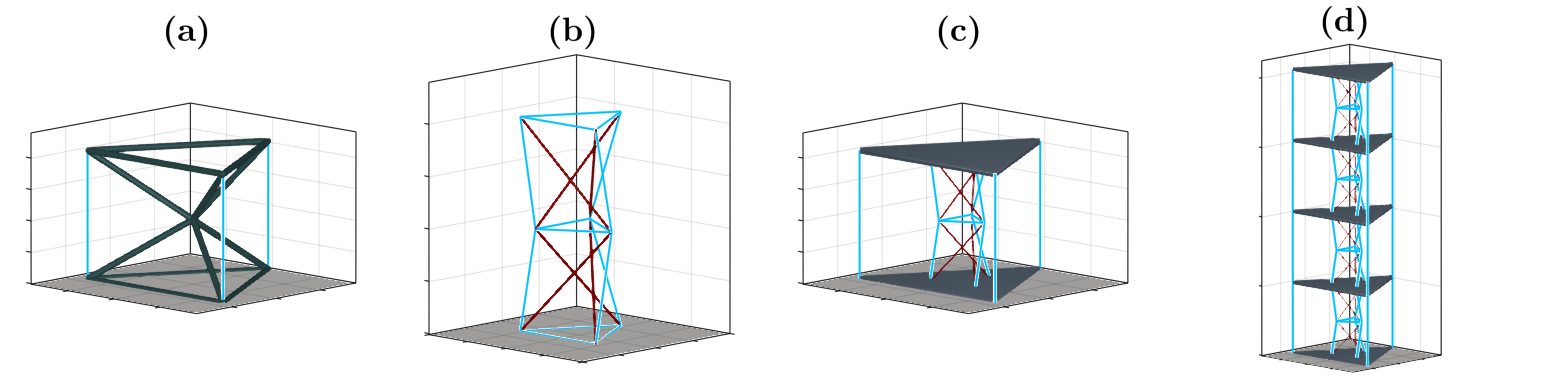}
	\caption{
			Schematic figures of (a,b) the primitive tensegrities, 
			(c) 1-stage, and (d) 4-stage \enquote{embedded} tensegrity structures.
	}
	\label{fig:embedding}
\end{figure}
\begin{figure}[tbh]
	\centering
	\includegraphics[width=\lw]{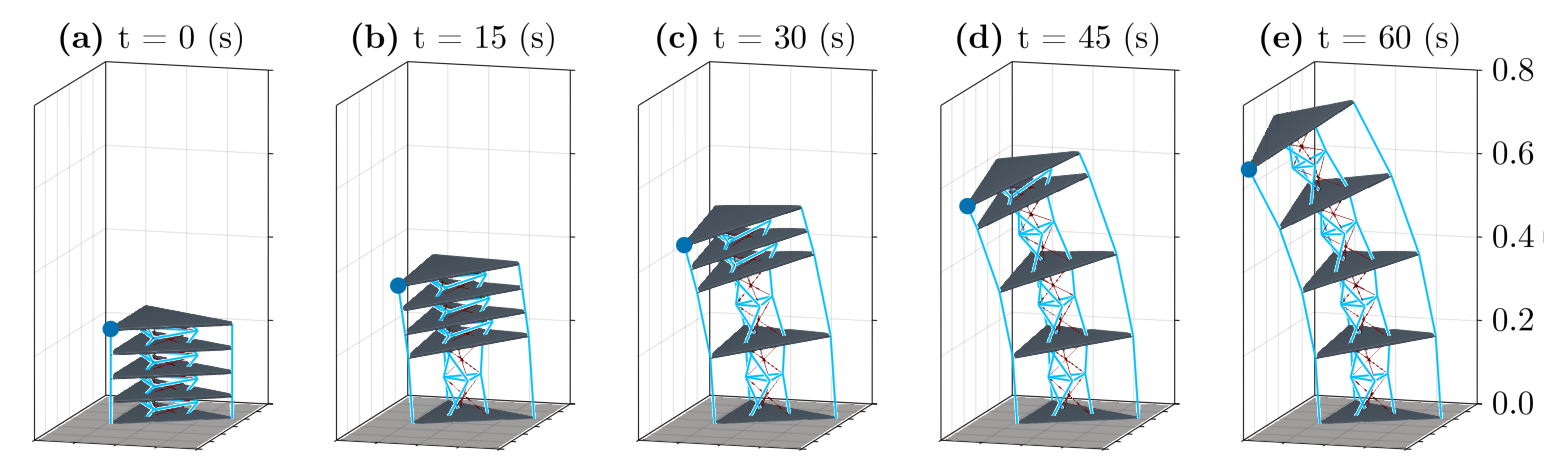}
	\caption{
		Snapshots of the 4-stage \enquote{embedded} tensegrity structures during cable-based deployment.
		Blue dots indicate the marker point. 
	}
	\label{fig:newembed0outerdeploy}
\end{figure}
\begin{figure}[th!]
	\centering
	\includegraphics[width=0.9\lw]{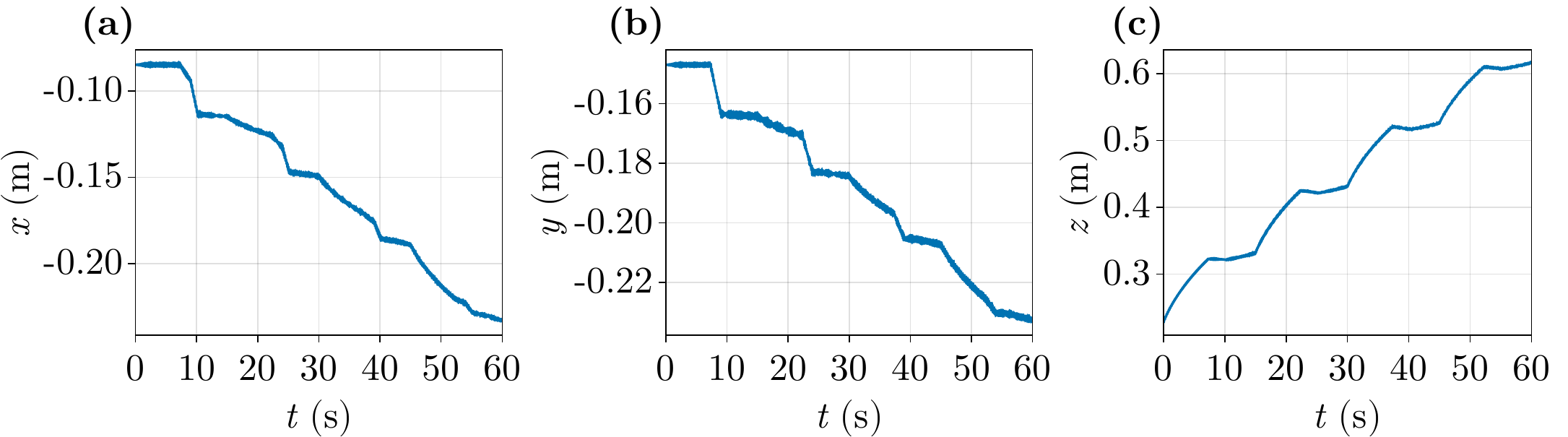}
	\caption{
		Trajectories of the marker point in (a) $x$, (b) $y$, and (c) $z$ directions.
	}
	\label{fig:newembed0outerdeploytraj}
\end{figure}

Besides of using algorithmic methods such as topology-finding, 
intuitive methods are also viable to design general tensegrity structures.
One such method can be called \enquote{embedding} as exemplified by \cref{fig:embedding}.
Firstly, the design process starts with known primitive tensegrities, 
such as a rotatable Class-2 tensegrity with two tetrahedrons in contact (\cref{fig:embedding} (a)), 
and a deployable 2-stage tensegrity prism (\cref{fig:embedding} (b)).
Secondly, the latter one can be embedded into the former one, replacing the ball joint  (\cref{fig:embedding} (c)).
Lastly, multiple modules can be stacked sequentially to build a multi-stage structures (\cref{fig:embedding} (d)).
In this way, the new structure is a Class-3 tensegrity 
endowed with the rotatable and deployable functionalities of the primitives.

Note that the \enquote{embedding} method is akin to 
the concept of \enquote{self-similar} iterations 
(See for example Ref.  \cite{skeltonTensegritySystems2009}), 
but not limited to \enquote{bars-only} compressive tensegrity structures.
Further in-depth investigations are still needed 
to broaden the applications of the \enquote{embedding} method, 
but those are beyond the scope of this paper.

Here, the new structure's  dynamic behaviors of rotations and deployments are demonstrated.
The structural properties are as follows. Each rigid bar has a length $l = \SI{0.14}{\meter}$ and 
a virtual radius of cross-section $r = \SI[exponent-mode = engineering]{0.0011667}{\meter}$, 
and the triangular rigid plate has a side length $l = \SI{0.2939394143492839}{\meter}$ and 
a height $h = \SI{0.01}{\meter}$. 
All rigid members are given a uniform density of teak wood $\rho = \SI{630}{\kilogram\per\meter^3}$,
and tensile cables are given a stiffness coefficient $\kappa = \SI{25.918}{\kilo\newton\per\meter}$.
Furthermore, to support the structure's self-weight, 
the lowermost plate is fixed to the ground by giving boundary conditions.

Deployments of the structure are achieved by cable-based actuation 
\cite{kanNonlinearDynamicDeployment2018a,roffmanCableActuatedArticulatedCylindrical2019},
which is implemented by changing the rest lengths of the cables.
For demonstration purpose, the outer three cables in each stage are given different rest lengths, 
while the cables in the prisms are released in a stage-by-stage manner. 
Simulation results are visualized in \cref{fig:newembed0outerdeploy}, 
showing that the structure can be successfully deployed in $\SI{60}{\second}$, 
while the uneven tensions of the outer cables lead to inclination. 
\cref{fig:newembed0outerdeploytraj} plots the trajectories of marker point in the deployment process, 
showing that small vibrations occur during the dynamic deployment due to rigid-tensile coupling.
The reduction of such vibrations is subject to further research.

\subsection{Example 4: A tensegrity structure designed by interfacing}

\begin{figure}[th!]
	\centering
	\includegraphics[width=0.7\lw]{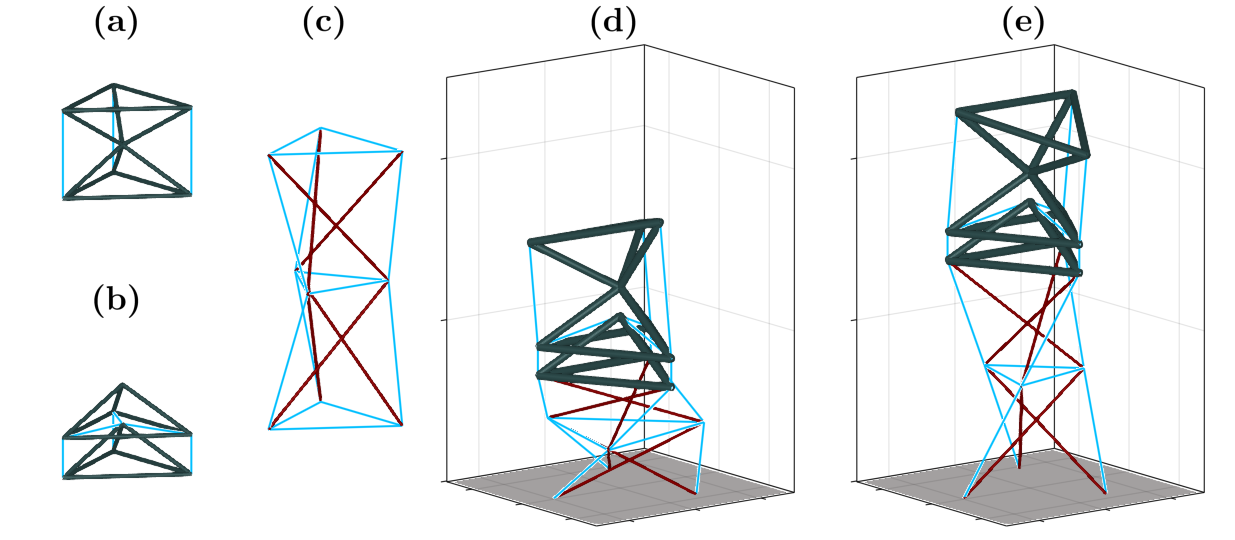}
	\caption{
		Schematic figures of (a,b,c) the primitive tensegrities, 
		the (d) initial and (e) target configurations of the tower-like tensegrity structure designed by interfacing. 
	}
	\label{fig:tower3d}
\end{figure}

\begin{figure}[th!]
	\centering
	\includegraphics[width=0.6\lw]{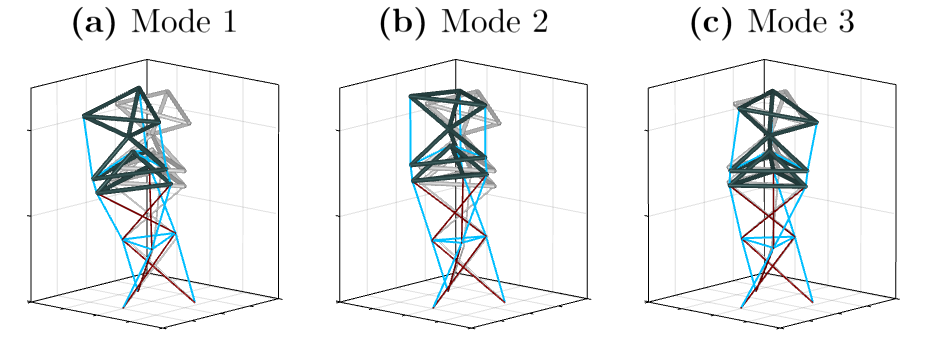}
	\caption{
		The mode shapes of the first three vibration modes of the tower-like 
        tensegrity structure in target configuration.
		The configuration of static equilibrium is colored in gray for reference.
	}
	\label{fig:tower3dmodes}
\end{figure}

\begin{figure}[th!]
	\centering
	\includegraphics[width=0.8\lw]{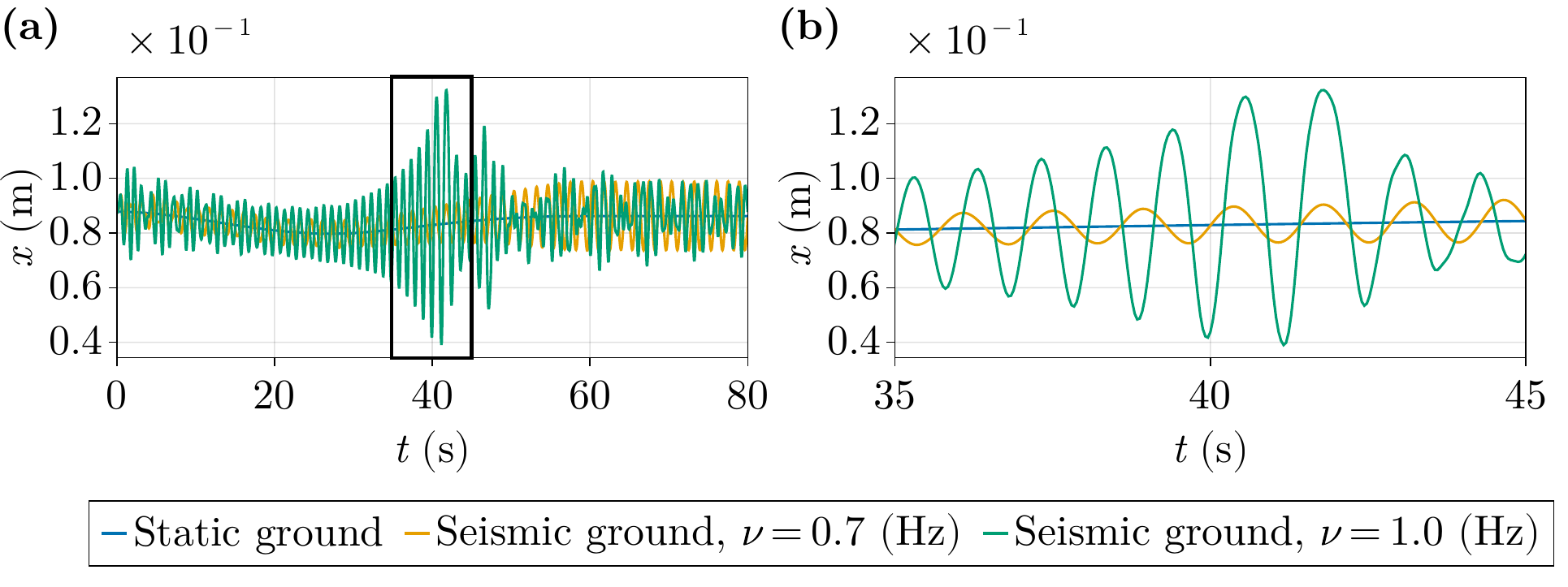}
	\caption{
		Trajectories of the marker point in the (a) $x$-direction and 
        (b) the enlargement view for the three deployment cases.
	}
	\label{fig:tower3ddpltraj}
\end{figure}

\begin{figure}[th!]
	\centering
	\includegraphics[width=0.8\lw]{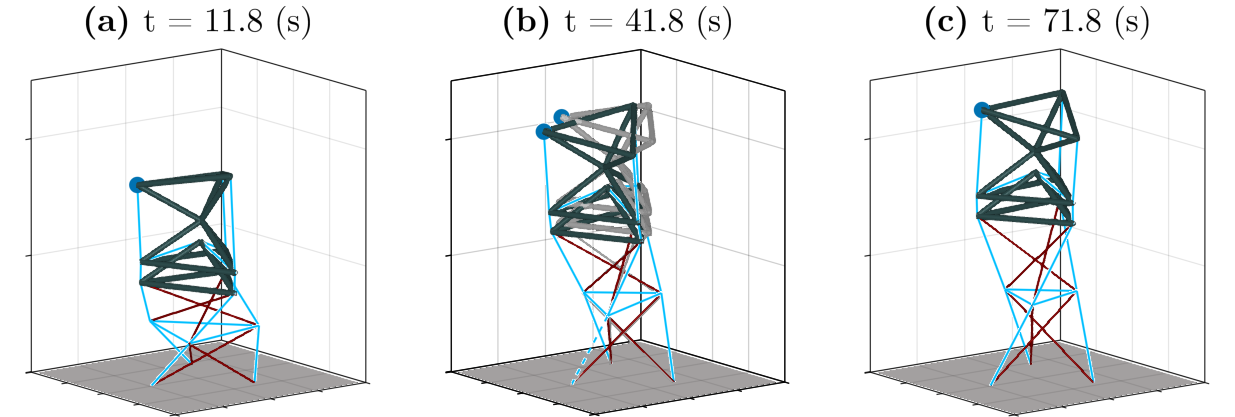}
	\caption{
		Snapshots of the tower-like tensegrity structure during deployment with seismic frequency $\nu=\SI{1.0}{\hertz}$.
		Dash blue lines indicate slack cables. 
        The deployment on static ground is colored in gray for reference in (b).
		Blue dots indicate the marker point. 
	}
	\label{fig:tower3dseis10}
\end{figure}
Another intuitive design methods can be called \enquote{interfacing}, as exemplified by \cref{fig:tower3d}.
Consider again the two tensegrity primitives in example 4, as shown in \cref{fig:tower3d} (a,c).
Additionally, a spine-like tensegrity primitive \cite{tietzTetraspineRobustTerrain2013}, composed of 2 tetrahedrons and  6 cables, 
are introduced as an interface to connect the former two, 
leading to a new multi-stage tower-like structure \cref{fig:tower3d} (d,e).
In this way, the new structure also acquires the ability of rotations and deployments, 
albeit at different stages. 
The advantage of the \enquote{interfacing} method is that 
it can extend an existing structure,  
without altering its internal configuration.
Thus, it automatically leads to modular structure designs, 
and can be easily combined with other methods, 
such as topology-finding and the \enquote{embedding} method.

Here, we simulate the resonance phenomenon during deployment with moving boundary conditions to validate the dynamic formulations and the MSI scheme.
The structural properties are as follows. 
Each rigid bar has a length $l = \SI{0.22}{\meter}$, 
a virtual radius of cross-section $r = \SI[exponent-mode = engineering]{0.00183333}{\meter}$,  
and a uniform density $\rho = \SI{630}{\kilogram\per\meter^3}$.
Each tetrahedron has a height $h = \SI{0.0707107}{\meter}$, 
a circumradius $r = \SI{0.1}{\meter}$ for the base triangle, 
with mass $m = \SI{0.299923}{\kilogram} $
and inertia matrix $\bar{\bm{I}}= 
\mathrm{diag}\left(
 \num{0.76639282053},
 \num{0.76638139752},
 \num{1.2464720496}
 \right)
 \times 10^{-3}
 \unit{\kilogram\square\meter} 
 $.
Tensile cables in the prism are given the stiffness coefficient $\kappa = \SI{1e3}{\newton\per\meter}$.
Otherwise $\kappa = \SI{5e2}{\newton\per\meter}$.
All tensile cables have the same damping coefficient $\eta = \SI{2}{\newton\per\meter\per\second}$.
The lower ends of the tensegrity prism are fixed to the ground, 
which is subject to a seismic wave in the $x$ direction $x(t) = 0.003 \sin(\nu 2\pi t) \unit{\meter}$, 
where $\nu$ is the seismic frequency.

Firstly, two static equilibrium configurations (\cref{fig:tower3d}(c,d)) with 
different cable rest lengths are sought using the geodesic dynamic relaxation method 
\cite{mikiGeodesicDynamicRelaxation2014}.
These two static equilibria will be referred to as the initial and target states 
for the deployment process.
According to linearized dynamic analysis, 
the lowest natural frequencies for these two states are 
$\xi = \SI{1.254728}{\hertz}$ and $\xi = \SI{0.82885849}{\hertz}$, respectively.
The first three vibration modes of the target state are shown in \cref{fig:tower3dmodes}.
Based on these preliminary results, it is expected that resonances would occur during 
deployment if the seismic frequency $\nu$ is within the range 
$\left[\num{0.8288584}, \num{1.254728}\right]\unit{\hertz}$.
To verify this prediction, cable-based deployment simulations are carried out with 
three different seismic frequencies $\nu = 0, 0.7, 1.0~\unit{\hertz}$.
An 80-second simulation with 60-second deployment time is carried out.
Trajectories of the marker point are plotted in \cref{fig:tower3ddpltraj}.
It shows that the amplitude of response in the $x$ direction is 
significantly increased only for $\nu = \SI{1.0}{\hertz}$, verifying our prediction.
In fact, the resonant vibrations are large enough to cause cable-slacking, 
as shown in \cref{fig:tower3dseis10} (b).
To sum up, these results validate the proposed approach and demonstrate 
its efficacy in dealing with complex conditions, 
including slack cables, cable-based deployment, and moving boundary conditions.

\section{Conclusion}\label{sec:conclusion}
In this paper, a unified approach is developed for dynamic analysis of general tensegrity structures.
The heterogeneity between 6-DoF rigid bodies and 5-DoF rigid bars is resolved by the non-minimal description of natural coordinates.
And the resulting dynamic equation has a constant mass matrix and is free from trigonometric functions.

The natural coordinates are comprehensively reformed.
Four and two types of natural coordinates are derived for a 3D rigid body and a rigid bar.
The idea of polymorphism unifies different types of coordinates, 
and thereby facilitates the formulations for ball joints, boundary conditions, 
and cables' tension forces for general tensegrity structures.

To enable modal analysis, the governing DAEs is linearized 
around static equilibrium and then reduced to a linear system 
using the reduced-basis method.

To enable numerical simulation of nonlinear dynamics, 
central difference approximations and 
the technique of taking limits at time-segments' endpoints are 
used to derive a one-timestep modified symplectic integration (MSI) scheme 
that not only yields realistic results for energy and vibrations in long-time 
simulations, but also accommodate non-conservative forces and boundary conditions.

Four representative numerical examples are presented.
Examples 1 and 2 are general tensegrity found in the topology-finding literature, 
while examples 3 and 4 are novel multi-functional structures created by two intuitive ways, 
namely the \enquote{embedding} and \enquote{interfacing} methods.
Various complex situations, including dynamic external loads, 
cable-based deployment, and moving boundaries, 
demonstrate the efficacy of the proposed approach for 
the dynamic analysis of general tensegrity structures.

Regarding future research, the proposed approach can be extended to include cable masses 
\cite{goyalTensegritySystemDynamics2019,kanNonlinearDynamicDeployment2018a}, 
which are ignored in this paper, and clustered cables, 
which slide through pulleys on the rigid members to achieve clustered actuation 
\cite{kanComprehensiveFrameworkMultibody2021,maDynamicsControlClustered2022,belhadjaliStaticAnalysisTensile2021} . 
These extensions should be straightforward due to the same underlying framework of Lagrangian mechanics.
Finally, the linear dependence on cables' force densities \cref{eq:tension_force} 
can be exploited for the design of control schemes \cite{chenDesignControlTensegrity2020}, 
aiming to integrate structure and control design as 
for classical tensegrity systems \cite{goyalIntegratingStructureInformation2021}.

\backmatter


\section*{Declarations}
\bmhead{Funding}
This research was supported by the National Natural Science Foundation of China (Grant numbers: 12002396, 12232015, 11972102) 
and Shenzhen Science and Technology Program (Grant number: ZDSYS20210623091808026).
\bmhead{Competing interests}
The authors have no competing interests to declare that are relevant to the content of this article.









\bibliography{DynamicTensegrityPaper}

\end{document}